\documentclass[preprint]{aastex}

\newcommand       \be           {\begin{equation}}
\newcommand       \ee           {\end{equation}}
\newcommand       \Angstrom     {\,{\rm \AA}}          
\newcommand       \eV           {\,{\rm eV}\,}
\newcommand       \K            {\,{\rm K}}
\newcommand       \cm           {\,{\rm cm}}
\newcommand       \erg          {\,{\rm erg}}
\newcommand	  \g		{\,{\rm g}}

\newcommand	  \pc		{\,{\rm pc}}

\newcommand       \nH           {n_{\rm H}}
\newcommand	  \s		{\,{\rm s}}
\newcommand       \Tc           {T_{\rm c}}
\newcommand       \urad         {u_{\rm rad}}
\newcommand       \uHab         {u_{\rm Hab}^{\rm uv}}
\newcommand       \Jpe          {J_{\rm pe}}
\newcommand       \Jion         {J_{\rm ion}}
\newcommand	  \pdt		{{\rm pdt}}
\newcommand	  \pet		{{\rm pet}}
\newcommand       \Qabs	        {Q_{\rm abs}}

\newcommand    	  \Gampe        {{\Gamma}_{\rm pe}}
\newcommand	  \Lamgr	{{\Lambda}_{\rm gr}}

\newcommand       \gtsim        {\gtrsim}
\newcommand       \ltsim        {\lesssim}
\newcommand	  \E	        {\erg \cm^{-3}}

\shorttitle{Photoelectric Emission from Dust}
\shortauthors{Weingartner \& Draine}

\begin{document}

\title{Photoelectric Emission from Interstellar Dust:  Grain Charging
and Gas Heating}

\author{Joseph C. Weingartner}
\affil{Physics Dept., Jadwin Hall, Princeton University,
        Princeton, NJ 08544, USA; CITA, 60 St. George Street, University of
Toronto, Toronto, ON M5S 3H8, Canada}
\email{ weingart@cita.utoronto.ca}

\and

\author{B.T. Draine}
\affil{Princeton University Observatory, Peyton Hall,
        Princeton, NJ 08544, USA} 
\email{draine@astro.princeton.edu}

\begin{abstract}

We model the photoelectric emission from and charging of
interstellar dust and
obtain photoelectric gas heating efficiencies as a function of grain size
and the relevant ambient conditions.
Using realistic grain size distributions, we
evaluate the net gas heating rate for various interstellar environments, 
and find less heating for dense regions
characterized by $R_V = 5.5$ than for diffuse regions with $R_V = 3.1$.
We provide fitting functions which reproduce our numerical results for
photoelectric heating and recombination cooling for a wide range of 
interstellar conditions. 
In a separate paper we will examine the implications of these results for
the thermal structure of the interstellar medium.
Finally, we investigate the potential importance of photoelectric heating in
H~II regions, including the warm ionized medium.
We find that photoelectric heating could be comparable to or
exceed heating due to photoionization of H for high ratios of the
radiation intensity to the gas density.  We also find that
photoelectric heating by dust can account for the observed variation of 
temperature with distance from the galactic midplane in the warm ionized
medium.  

\end{abstract}

\keywords{dust --- extinction --- HII regions --- ISM: clouds}

\section{Introduction}

Photoelectric emission from dust grains dominates the heating of diffuse 
interstellar gas clouds as well as the photodissociation regions at the 
surfaces of molecular clouds.  This mechanism therefore plays an important
role in the dynamical evolution of the interstellar medium, and
in shaping the line emission spectra of galaxies.
Heating by photoelectric emission from interstellar grains was first
considered by Spitzer (1948), and there have since been a number of
reassessments of the heating rate
(Watson 1972; deJong 1977; Draine 1978; 
Tielens \& Hollenbach 1985; Bakes \& Tielens 1994, Dwek \& Smith 1996).

For a given grain, the heating rate depends on the grain size, composition,
and charge state, as well as on the spectrum of the illuminating radiation.
Because photoelectric yields are enhanced for small grains (Watson 1972, 1973),
estimates of the net photoelectric heating rate in interstellar gas are
sensitive to the adopted grain size distribution, which
should be consistent with the observed extinction curve
(which shows strong regional variations) as well as with the observed
emission from interstellar dust grains, 
from the near-infrared to the microwave.
Previous estimates for photoelectric heating rates did not always
use grain size distributions which were consistent with these constraints.
While the measured extinction curve by itself does not suffice to uniquely
specify the grain size distribution, in the present study we will
consider size distributions which are consistent with the observed
extinction in different regions, with either the minimum or maximum
permissible population of ultrasmall grains (Weingartner \& Draine 2000).


In addition to using size distributions consistent with observations,
we also model the photoelectric emission process and associated
grain charging in detail, using realistic yields for graphitic and
silicate grains, 
a realistic distribution of photoelectron kinetic energies,
and new estimates for electron sticking efficiencies for small grains.
The resulting photoelectric heating rates are calculated for grain
size distributions consistent with extinction curves characteristic
of diffuse clouds ($R_V\equiv A_V/E_{B-V} \approx 3.1$),
intermediate density regions ($R_V \approx 4.0$),
and dense clouds ($R_V \approx 5.5$).

In \S \ref{sec:peemission} 
we discuss our adopted model for photoelectric emission from grains, 
expanding on the treatments of Draine (1978) and Bakes \& Tielens (1994).
In \S \ref{sec:sticking} we use recent experimental results to
obtain improved estimates for the electron sticking coefficient,
as a function of grain size; our new sticking coefficients differ
significantly from previous estimates.
In \S \ref{sec:ambient} 
we characterize the relevant ambient conditions (which are important
in grain charging) and in
\S \ref{sec:peheateff} 
we evaluate the photoelectric heating efficiency as a function of 
grain size.  
We obtain net heating rates in H~I regions 
in \S \ref{sec:netheat}
and in \S \ref{sec:HIIregions} we
investigate the potential contribution of photoelectric 
heating in H~II regions, finding that grain photoelectric heating may
explain observed high temperatures in the warm ionized medium.
We briefly summarize our results in \S \ref{sec:summary}.

\section{Photoelectric Emission from Grains\label{sec:peemission}}

\subsection{Grain Characterization}

We consider both graphitic and silicate grains.  We adopt the description 
of graphitic grains given by Li \& Draine (2000), in which the smallest
grains are polycyclic aromatic hydrocarbon (PAH) molecules, the largest
grains consist of graphite, and grains of intermediate size have optical 
properties intermediate between those of PAHs and graphite.  We
characterize the graphitic grains by an effective radius $a$, the radius of
a sphere containing the same number of C atoms,
in which C atoms contribute a mass density $\rho = 2.24 \g \cm^{-3}$,
the density of ideal graphite.  The smallest grains certainly are not 
spherical and probably are not well characterized by the ideal graphite 
density; in this case the effective radius simply indicates the number 
$N_{\rm C}$ of C atoms in the grain:
\be
\label{eq:nc-a}
N_{\rm C} = 468 \left(\frac{a}{10^{-7}\cm}\right)^3 ~~.
\ee
For example, coronene C$_{24}$H$_{12}$ is assigned an effective radius
$a=3.7\Angstrom$, and a surface area $4\pi a^2 = 173\Angstrom^2$.

In following sections, the $N_{\rm C}$--$a$ relation (\ref{eq:nc-a})
will be used to incorporate experimental data on hydrocarbon
molecules into our description in terms of the effective radius.  
Since small molecules
that extrapolate to silicates as $a \rightarrow \infty$ have not been
studied, we will not need a similar relation for silicates.  In this case,
$a$ is simply the radius of a spherical silicate grain.

\subsection{Electron Energy Levels}

\begin{figure}
\epsscale{1.00}
\plotone{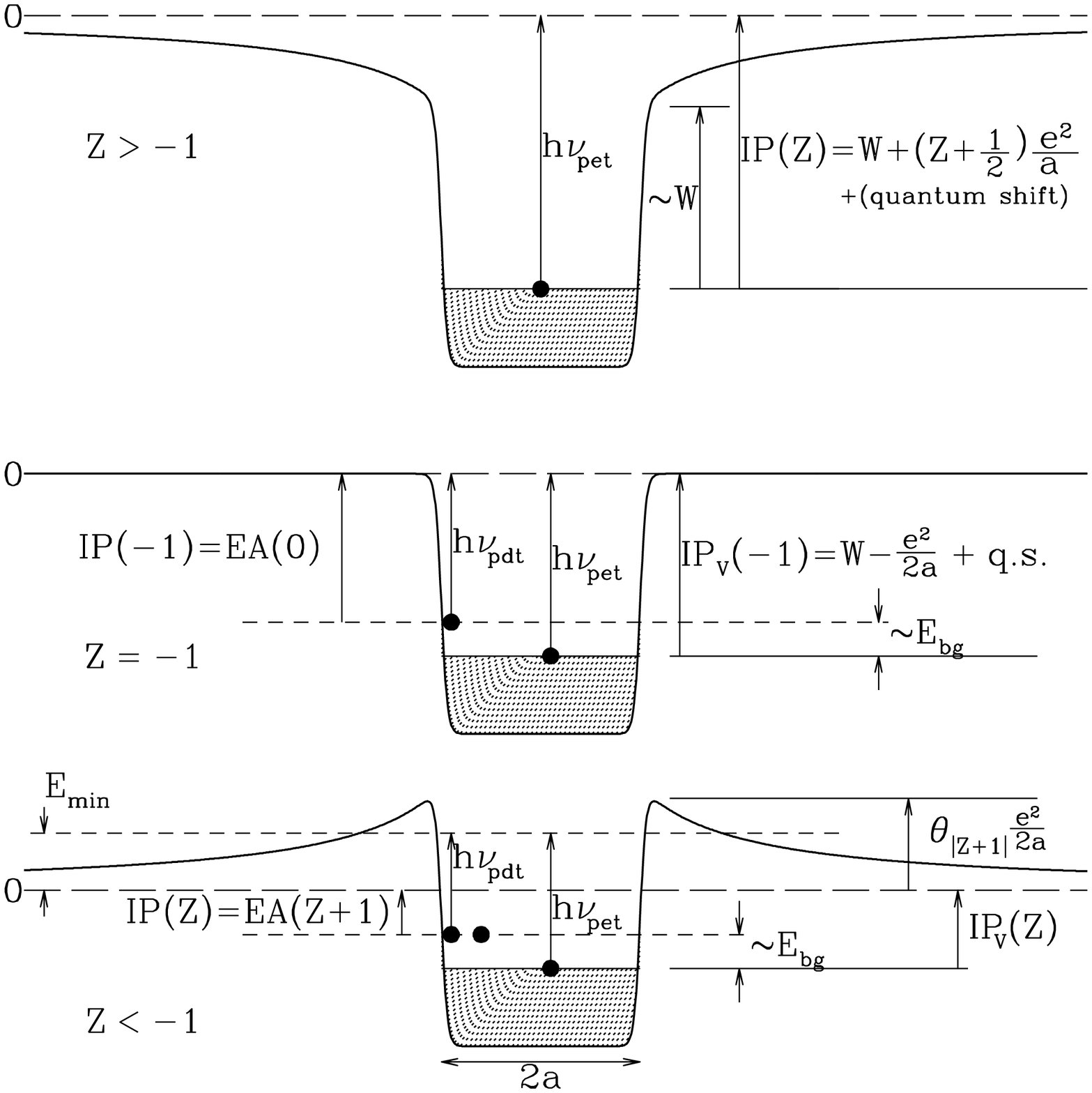}
\caption{
\label{fig:pot_wells}
The potential confining electrons in a grain with charge $Ze$.  Shaded 
regions show occupied energy levels and $W$ is the work function.
Upper panel:  $Z>-1$.  The photoemission threshold photon energy
$h\nu_\pet = IP$, the ionization potential.
Middle panel:  $Z=-1$.  The extra electron occupies the lowest unoccupied
energy level (LUMO) of the neutral grain, lying a distance $EA(Z=0)$ below
zero (EA is the electron affinity).  The ionization potential $IP(Z=-1)$
and photodetachment threshold energy $h\nu_{\rm pdt}$
both equal $EA(Z=0)$.  The photoemission 
threshold energy $h\nu_\pet = IP_V$, the valence band ionization potential.  
$IP_V - IP \approx E_{\rm bg}$, the energy band gap in bulk material; the 
equality is not exact due to quantum shifts in energy levels with grain size.
Lower panel:  $Z<-1$.  When an electron acquires an energy $E_{\rm min}$
above zero, it can tunnel out of the grain.  Thus, $h\nu_{\rm pdt} = 
IP(Z)+E_{\rm min}$ and $h\nu_{\rm pet} = IP_V(Z) + E_{\rm min}$.  When 
$Z<-1$, the maximum value of the confining potential is 
$\theta_{|Z+1|} e^2/a$; the parameter $\theta_{|Z+1|}$ is derived by 
Draine \& Sutin (1987).
	}
\end{figure}

The electrons in a solid are confined in a potential well 
(see, e.g., Lang \& Kohn 1971).
For neutral bulk material, the work function $W$ is the energy 
of a free electron with zero kinetic energy minus the energy of an electron
in the highest occupied state in the solid. (Throughout this section, refer
to Figure \ref{fig:pot_wells} for schematic depictions of the
electron-confining potential.)  In this 
case, the ionization potential $IP$, i.e., the difference in energy between
infinity and the highest occupied state, equals $W$. 

For small grains, $IP \neq W$.  We first consider grains which are not so 
small that quantum effects shift the energy levels from their bulk values.
If $Z \ge 0$, then an electron that is liberated from the grain material
feels a Coulomb attraction to the grain, which now has charge $(Z+1) e$.  
For a conducting sphere, $IP-W$ can be determined by calculating the work 
needed to remove the electron from a point just outside the sphere to infinity
(including the term due to the image charge as well as that due to the net
charge on the sphere) and subtracting the corresponding work associated with
a plane surface.  The result is that $IP = W + (Z+1/2)e^2/a$ (see, e.g. 
Makov, Nitzan, \& Brus 1988; Seidl \& Brack
1996).\footnote{Surprisingly, there has been some
controversy in the literature as to the value of the constant added to $Z$.
See Makov et al.\ (1988), Moskovits (1991),
Seidl \& Brack (1996).}  For nonconducting materials,
the term $e^2/2a$ above should be multiplied by $(\epsilon -1)/\epsilon$, 
where $\epsilon$ is the dielectric function (Makov et al.\ 1988).  This 
correction is fairly small for candidate grain materials, so we will ignore 
it.  

When $Z<0$, the ``extra'' electrons, which occupy the ``LUMO''
(for ``lowest unoccupied molecular orbital'') of the neutral, are separated
from the top of the valence band by the band gap, with extra energy 
$E_{\rm bg}$ (assuming the valence band is full in the neutral).  
The electron affinity $EA(Z)$ is the difference in energy
between infinity and the LUMO for the grain of charge $(Z-1) e$ created by
adding the electron.  Thus, $IP(Z<0) = EA(Z+1)$.  We also define the
``valence band ionization potential'' $IP_V(Z)$, equal to the difference 
in energy between infinity and the top of the valence band.  Of course,
$IP_V(Z)=IP(Z)$ when $Z\ge 0$.  

\begin{figure}
\epsscale{1.00}
\plotone{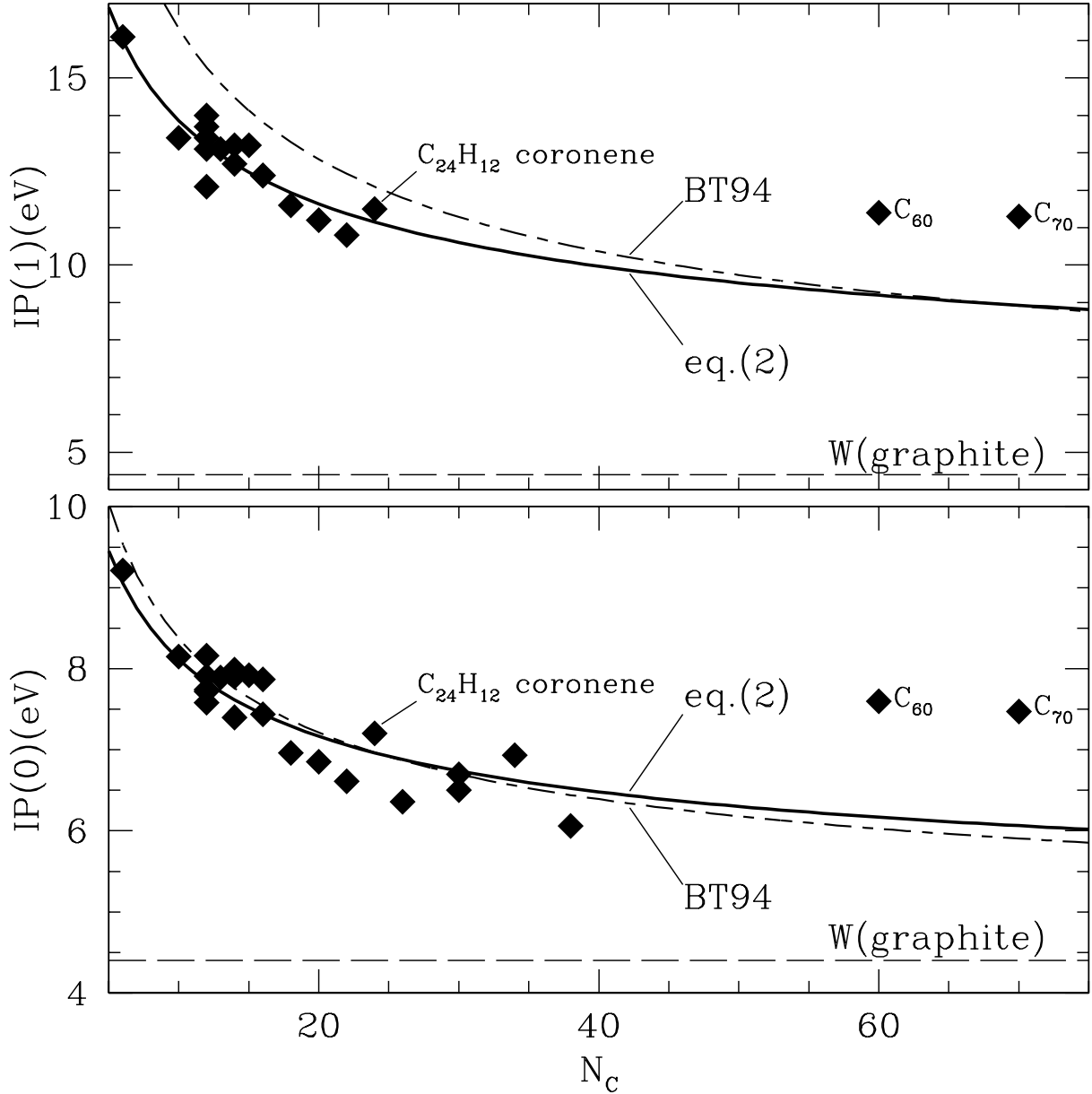}
\caption{
\label{fig:IP}
	First and second ionization potentials 
	$IP(0)$ and $IP(1)$ for
	cyclic aromatic hydrocarbons and fullerenes.
	Solid curve is eq.(\ref{eq:IP}).
	Curve labelled BT94 is the $IP$ estimate
	of Bakes \& Tielens (1994).
	Data:
	Lias et al.\ (1988) for 
	C$_{14}$H$_{10}$ anthracene, 
	C$_{26}$H$_{16}$ hexacene,
	C$_{30}$H$_{14}$ dibenz[bc,hl]-coronene,
	C$_{30}$H$_{16}$ pyranthene,
	C$_{34}$H$_{18}$ tetrabenz[a,cdj,lm]-perylene;
	Tobita et al.\ (1994) for 
	C$_{10}$H$_{8}$ naphthalene,
	C$_{12}$H$_{8}$ biphenylene,
	C$_{12}$H$_{8}$ acenaphthlylene,
	C$_{12}$H$_{10}$ biphenyl,
	C$_{12}$H$_{10}$ acenaphthene,
	C$_{12}$H$_{10}$ 2-vinylnaphthalene,
	C$_{13}$H$_{10}$ fluorene,
	C$_{14}$H$_{10}$ diphenylacetylene,
	C$_{14}$H$_{12}$ 9,10-dihydrophenanthrene,
	C$_{14}$H$_{9}$N acridine,
	C$_{16}$H$_{10}$ pyrene
	C$_{16}$H$_{10}$ fluoranthene,
	C$_{18}$H$_{12}$ tetracene,
	C$_{20}$H$_{12}$ perylene,
	C$_{22}$H$_{14}$ pentacene,
	C$_{24}$H$_{12}$ coronene;
	Chen et al.\ (1999) for C$_{38}$H$_{20}$ benz[42];
	Lichtenberger et al.\ (1991),
	Lichtenberger et al.\ (1992), and Steger et al.\ (1992) for 
	C$_{60}$, C$_{70}$ fullerenes.
	}
\end{figure}

Since it is not yet possible to calculate the quantum shifts in energy 
levels for very small grains, we adopt an empirical approach.    
Figure \ref{fig:IP} shows the first and second ionization potentials
$IP(0)$ and $IP(1)$ for various aromatic hydrocarbons,
including C$_6$H$_6$ benzene, C$_{24}$H$_{12}$ coronene,
and C$_{38}$H$_{20}$ benz[42].
The polycyclic aromatic hydrocarbons (PAHs) of interest here 
appear to be fit by 
\be
IP_V(Z) = W + \left(Z+\frac{1}{2}\right) \frac{e^2}{a}
+ (Z+2) \frac{e^2}{a} \left(\frac{0.3\Angstrom}{a}\right) ~~~,
\label{eq:IP}
\ee
with the graphite work function $W=4.4\eV$.
This estimate for $IP(Z)$ is plotted in Figure \ref{fig:IP} for
$Z=0$ and 1.  For the first ionization potential $IP(0)$, equation 
(\ref{eq:IP}) is numerically very close to the estimates of
Bakes (1992) and Bakes and Tielens (1994), but for $IP(1)$ our
estimate is $\sim$1~eV below their estimate for PAHs with $\sim$15-30
C atoms.  Note that equation (\ref{eq:IP}) reproduces the classical 
electrostatic result for large $a$.  Since experimental evidence is not
available for other ionization states, we will adopt equation
(\ref{eq:IP}) for all values of $Z$.  

For silicates, few laboratory experiments are available to guide us.  We
adopt $W = 8 \eV$, since the photoelectric yield for lunar dust drops 
steeply around that energy (Feuerbacher et al.\ 1972).  Nayak et al.\ (1998)
have made theoretical calculations of the ionization potentials of small 
silica clusters (SiO$_2$)$_n$, with $n=$ 1 to 6.  They find $IP(0)=11.19 \eV$
($16.08 \eV$) for $n=2$ (5), and intermediate results for other values of 
$n$.  If we employ equation (\ref{eq:IP}) with $W=8 \eV$ and $a \approx 3.2 
\Angstrom$, then we find $IP(0) \approx 12\eV$, which is in the Nayak et al.\ 
range for small clusters.  Thus, we adopt equation (\ref{eq:IP}) for both 
carbonaceous and silicate grains, with $W=4.4$ and $8\eV$, respectively.

We expect that
\be
EA(Z\leq0)=W-E_{\rm bg}(a)+\left(Z-\frac{1}{2}\right)
\frac{e^2}{a} + O(a^{-2}) ~~~,
\ee
where the presence of an $O(a^{-2})$ term is expected since such a
correction was required to fit the experimental results for $IP_V(Z\geq0)$
(eq. \ref{eq:IP}).
The ``infinite limit'' for PAHs -- graphite --
is a ``semimetal'', 
with a slight
overlap between the top of the valence band and the conduction band
(Spain 1973).
For metals and semimetals, $E_{\rm bg}=0$.  Many silicates begin to absorb
strongly around 2500--$2000 \Angstrom$, corresponding to a band gap of
5--$6 \eV$; we will adopt $E_{\rm bg} = 5 \eV$, independent of $a$, for 
silicates.

\begin{figure}
\epsscale{1.00}
\plotone{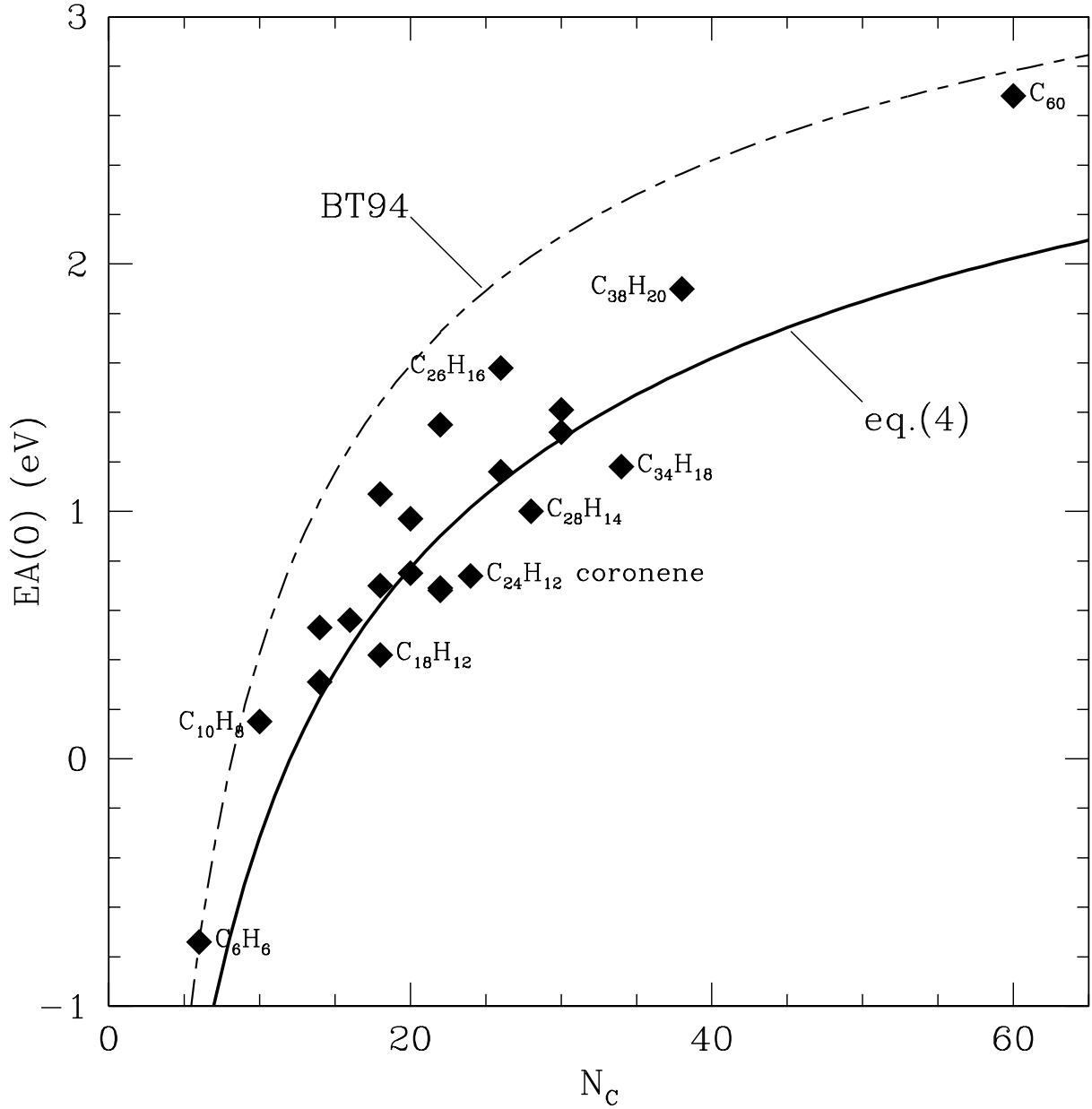}
\caption{
\label{fig:EA}
	Electron affinity for neutral PAH molecules and C$_{60}$.
	Solid curve is eq.(\ref{eq:EA}).
	Curve labelled BT94 is estimate of Bakes (1992), used by Bakes \&
	Tielens (1994).
	Data: Chen et al.\ (1999) for 
	C$_{6}$H$_{6}$ benzene,
	C$_{24}$H$_{12}$ coronene,
	C$_{26}$H$_{16}$ hexacene,
	C$_{30}$H$_{14}$ dibenz[bc,hl]-coronene,
	C$_{30}$H$_{16}$ pyranthene,
	C$_{34}$H$_{18}$ tetrabenz[a,cdj,lm]-perylene,
	C$_{38}$H$_{20}$ benz[42];
	Shiedt \& Weinkauf (1997) for 
	C$_{14}$H$_{10}$ anthracene;
	Ruoff et al.\ (1995) for
	C$_{10}$H$_{8}$ naphthalene,
	C$_{14}$H$_{10}$ phenanthrene,
	C$_{16}$H$_{10}$ pyrene,
	C$_{18}$H$_{12}$ tetracene,
	C$_{18}$H$_{12}$ benzanthracene,
	C$_{18}$H$_{12}$ chrysene,
	C$_{20}$H$_{12}$ perylene,
	C$_{20}$H$_{12}$ benzo(a)pyrene,
	C$_{22}$H$_{14}$ pentacene,
	C$_{22}$H$_{14}$ dibenz(a,j)anthracene;
	Chen et al.\ (1997) for
	C$_{26}$H$_{12}$ diindenochrysene,
	C$_{28}$H$_{14}$ debenzo[a,g]corannulene;
	and
	Wang, Ding, \& Wang (1999) for
	C$_{60}$ fullerene.
	}
\end{figure}

Values of $EA(0)$ are shown in Figure \ref{fig:EA} for selected neutral PAHs,
benzene, and C$_{60}$.
The relation
\be 
EA(Z) = W + \left( Z-\frac{1}{2}\right)\frac{e^2}{a} - \frac{e^2}{a}
\left(\frac{4\Angstrom}{a+7\Angstrom}\right) 
\label{eq:EA}
\ee
is also plotted in Figure \ref{fig:EA}, for $Z=0$ and $W=4.4\eV$.
For $a\rightarrow\infty$ this has the theoretically expected
behavior, and it provides a reasonable fit to
the experimental values of $EA(0)$ for benzene and the
PAHs in Figure \ref{fig:EA}
(for which $a\approx 2.3 - 4.3\Angstrom$).  We adopt this equation 
(\ref{eq:EA}) for carbonaceous grains.
For silicates, we adopt
\be
\label{eq:EAsil}
EA(Z) = W - E_{\rm bg} + \left( Z-\frac{1}{2}\right)\frac{e^2}{a},
\ee
with $W - E_{\rm bg} = 3 \eV$.  This yields $EA(0) \approx 0.73 \eV$ for
$a \approx 3.2 \Angstrom$, which is somewhat low compared with the values 
of 2--$3 \eV$ inferred for small silica clusters (Wang et al.\ 1997).

\subsection{Photoelectric Yields}

The photoelectric yield is the probability that an electron will be 
ejected following the absorption of a photon.  Calculation of the yield from
first principles is not yet possible.
Draine (1978) displayed yield measurements from 
experiments on bulk samples of materials of interest, including graphite, 
lunar dust, and silicon carbide.  
Watson (1972, 1973) pointed out that the
yields of submicron particles are 
expected to be enhanced relative to the bulk yields, because of the finite 
electron escape length.  An electron excited somewhere in the volume of the
sample can lose energy during its journey to the surface, through
interactions with electrons and with phonons.  The
result is that the fraction of electrons which escape energy loss
goes roughly as exp$(-x/l_e)$, where $x$ 
is the distance the electron has traveled and $l_e$ is the 
``electron escape length.''
Generally, the photon attenuation length, $l_a$ (equal to the 
$e$-folding length for the decrease in radiation intensity as it propagates 
into the material), exceeds $l_e$, and the electrons excited deep inside a 
bulk sample do not
reach the surface.  Small grain sizes limit the distance from
electron excitation to surface, so the yield is enhanced.  

A detailed and accurate model for the above geometrical effect is not yet
available.  A series of experiments (Schmidt-Ott
et al.\ 1980; Burtscher \& Schmidt-Ott 1982; Burtscher et al.\ 1984;
M\"{u}ller et al.\ 1988) on free silver spheres with radii between $27$ 
and $54 \Angstrom$ found yield enhancements in excess of expectations by 
factors of several.  A later theoretical effort (Faraci, Pennisi, \&
Margaritondo 1989) was able to reproduce the 
results, but only by assuming that the condition for an electron to escape,
when incident on the surface from within, depends only on its energy and 
not on its direction of motion.  
The estimates adopted here for size-dependent yields from grains are based on
Watson's (1973) simplified model for the geometrical effect and 
must be regarded as provisional.  Reliable calculations of
effects involving photoelectric emission 
from interstellar dust will not be possible
until experiments have been performed on submicron grains of appropriate 
composition.         

For negatively charged grains we
will distinguish between ``photoelectric'' ejection of electrons from the
valence band and ``photodetachment'' of excess ``attached'' electrons.

\subsubsection{Valence Band Electrons}

We assume that, when $Z>0$, the highest occupied
energy level is very close to the top of the valence band, since
the number of electrons which have been removed from the grain is small 
compared with the total number of electrons in the grain.  Thus, for 
$Z\ge 0$, the threshold photon energy for photoelectric emission is given
by $h\nu_{\rm pet}(Z) = IP_V(Z) = IP(Z)$.  
When $Z < -1$, $h\nu_{\rm pet}(Z) > IP(Z)$, because the electron has to 
overcome the repulsive Coulomb barrier.  Suppose the tunneling probability 
becomes significant when the electron energy exceeds the potential at 
infinity by $E_{\rm min}$. Then $h\nu_{\rm pet}(Z) = IP_V(Z) + E_{\rm min}$.
Thus, we take
\be
h\nu_\pet(Z,a) = \cases{IP_V(Z, a) &, $Z \ge -1$\cr
IP_V(Z,a) + E_{\rm min}(Z,a) &, $Z < -1$\cr}~~~.
\ee

We estimate $E_{\rm min}(Z,a)$ as the energy for which the tunneling 
probability (evaluated using the WKB approximation)
equals $10^{-3}$; the tunneling probability rapidly increases as
the excited electron energy increases above this value.  The
barrier consists of the Coulomb potential $-(Z+1)e^2/r$ and the image 
potential $-e^2 a^3 / 2 r^2 (r^2 - a^2)$, where $r$ is the distance from 
the center of the spherical grain.\footnote{Within a few $\Angstrom$ of the 
surface,
the image potential deviates from the classical expression, and ``saturates''
to a constant value (see, e.g., Lang \& Kohn 1971).  However, this saturated
portion of the potential is always classically allowed, and therefore does
not affect the WKB estimate of the tunneling probability.}  We adopt the
following expression for $E_{\min}$, which reproduces the results of the WKB 
calculation fairly well when $-(Z+1)e^2/a \gtsim 0.5 \eV$:
\be
E_{\min}(Z<0,a) = - \left( Z+1 \right)\frac{e^2}{a} \left[ 1+ \left(
\frac{27 \Angstrom}{a} \right)^{0.75} \right]^{-1}~~~.
\label{eq:Emin}
\ee
When $-(Z+1)e^2/a < 1 \eV$, equation (\ref{eq:Emin}) overestimates 
$E_{\min}$ as defined above, with increasing severity as $-(Z+1)e^2/a
\rightarrow 0$.  However, in this limit $E_{\min} \rightarrow 0$ as well,
so that this overestimate will not substantially affect the photoemission 
calculations.  

We adopt a very simple model for estimating the photoelectric
yield $Y(h\nu, Z, a)$.  
We first consider the yield $y_0^{\prime}$ of electrons that traverse the 
surface layer, emerge from the grain surface, and have enough energy to 
overcome the image potential (we will call these ``attempting'' electrons).
For $Z < 0$, every attempting electron will escape, but when $Z \ge 0$, 
some of these electrons have insufficient energy to escape to infinity
and instead fall back to the grain.  
We assume the following form for 
$y_0^{\prime}$:
\be
y_0^{\prime} = y_0 (\Theta) y_1(h\nu, a)~~~,
\ee
where the parameter $\Theta$ is given by
\be
\Theta = \cases{
h\nu - h\nu_\pet + (Z+1)e^2/a &, $Z \ge 0$\cr
h\nu - h\nu_\pet &, $Z < 0$\cr 
}
\ee
and the factor $y_1(h\nu, a)$ accounts for the size-dependent geometrical
yield enhancement discussed above; $y_1$ depends on $h\nu$ because the 
photon attenuation length $l_a$ does.

We assume a parabolic energy
distribution for the attempting electrons:\footnote{A parabolic form is
consistent with typical laboratory photoelectron distributions which drop to 
zero when $E=0$ and $E=h\nu - W$ and peak somewhere in between (Draine 1978).}
\be
\label{eq:parab}
f_E^0(E) = \frac{6 (E - E_{\rm low}) (E_{\rm high} -E)}
{(E_{\rm high} - E_{\rm low})^3}~~~,
\ee
where $f_E^0(E) dE$ gives the fraction of attempting electrons with 
energy (with respect to infinity) between $E$ and $E + dE$.  When $Z<0$,
$E_{\rm low} = E_{\rm min}$ (eq. \ref{eq:Emin}) and $E_{\rm high} = 
E_{\rm min} + h\nu - h\nu_\pet$; when $Z \ge 0$, $E_{\rm low} = -(Z+1) 
e^2/a$ and $E_{\rm high} = h\nu - h\nu_\pet$.  

Let $y_2$ be the fraction of attempting electrons which escape to infinity:
\be
\label{eq:y1}
y_2(h\nu, Z, a) = 
\cases{
\int_0^{E_{\rm high}} dE f_E^0(E) = 
E_{\rm high}^2 (E_{\rm high} - 3 E_{\rm low})/(E_{\rm high} -
E_{\rm low})^3~~~&, $Z\geq 0$\cr
1		&, $Z < 0$\cr
}
\ee

Our resulting expression for the photoelectric yield is
\be
Y(h\nu, Z, a) = y_2(h\nu, Z, a)
\min \left[ y_0(\Theta) y_1(a, h\nu), 1 \right]~~~.
\label{yeqn}
\ee

Draine (1978) found that the following estimate for $y_1$ reproduces 
Watson's (1973) detailed results based on Mie theory, to within $20 \%$:
\be
y_1 = {\left( \frac{\beta}{\alpha}\right)}^2
\frac{\alpha^2-2\alpha+2-2
\rm{exp}(-\alpha)}{\beta^2-2\beta+2-2\rm{exp}(-\beta)}~~~,
\label{fyeqn}
\ee
where $\beta = a/l_a$ and $\alpha = a/l_a + a/l_e$.  

Electron escape 
lengths for most materials reach a minimum in the vicinity of tens of $\eV$
and increase at higher and lower energies.  The data are meager for
electron energies
below $10 \eV$, but Martin et al.\ (1987) found, for thin carbon films, that
$l_e$ fell into a broad minimum of $6 \Angstrom$ around $40 \eV$ and rose to
$9 \Angstrom$ at $6 \eV$.  
McFeely et al.\ (1990) found, for SiO$_2$, that $l_e$ varied from $5.7$
to $6.8 \Angstrom$ for electron energies between $8$ and $20 \eV$, and
one might expect similar values for silicates.  
Bakes
\& Tielens (1994) assumed $l_e = 10 \Angstrom$, independent of energy, and
we use the same value here, for both graphite and silicates.

The photon attenuation length is given by 
\be
\label{eq:l_a}
l_a = \frac{\lambda}{4 \pi {\rm{Im}}(m)}~~~,
\ee
where $\lambda$ is the wavelength in vacuo and $m(\lambda)$ is the complex
refractive index.  Graphite is a highly anisotropic material, so that the
dielectric function is a tensor.  This tensor may be diagonalized by choosing 
Cartesian coordinates with two of the axes lying in the ``basal'' plane; 
the third axis, normal to the basal plane, is called the ``{\it c}-axis''.
For graphite, we take
\be
l_a^{-1} = \frac{4 \pi}{\lambda} 
\left[ \frac{2}{3} {\rm Im}(m_{\perp})
+ \frac{1}{3} {\rm Im}(m_{\parallel}) 
\right]~~~,
\ee
where $m_{\perp}$ and $m_{\parallel}$ are for the electric field perpendicular
and parallel to the {\it c}-axis, respectively. 
In computing $l_a$, we use dielectric functions from
Draine \& Lee (1984) and Laor \& Draine (1993), modified to remove a silicate
feature of crystalline origin that is not present in the observed 
interstellar extinction or polarization (see Weingartner \& Draine 2000).
Since PAH dielectric functions are not 
available, we use graphite dielectric functions for all carbonaceous grains.

\begin{figure}
\epsscale{1.00}
\plotone{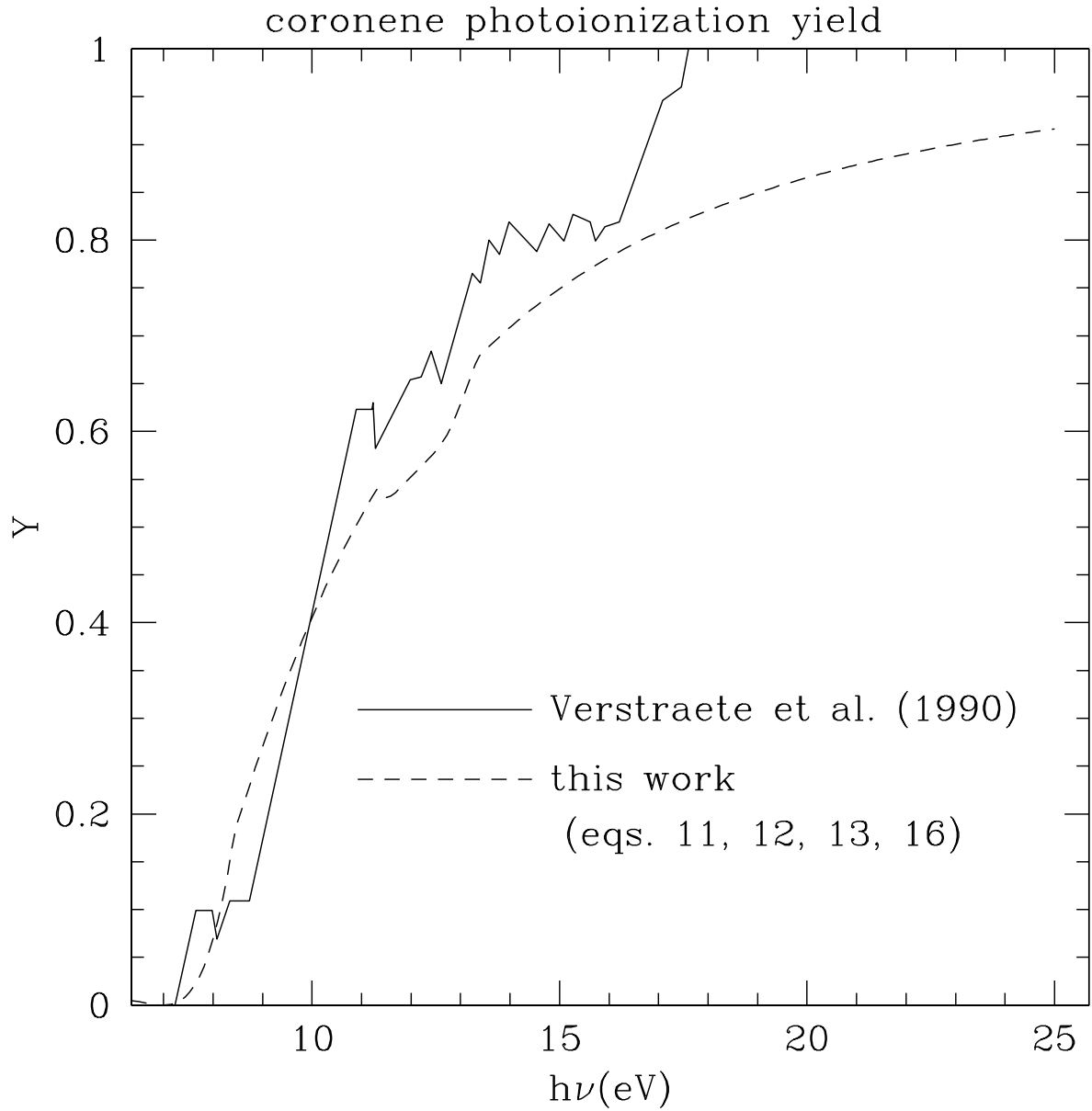}
\caption{
\label{fig:yield_coronene}
	Photoionization yield for coronene, as measured by Verstraete 
et al.~(1990) (solid) and as calculated using our model (dashed).
        }
\end{figure}

The final step in our yield prescription is to estimate $y_0(\Theta)$ for
carbonaceous and silicate grains.  
For carbonaceous grains, we follow Bakes \& Tielens (1994) and choose
$y_0$ such that $Y(h\nu, Z=0, a=3.7 \Angstrom)$
approximately reproduces the photoionization yield for coronene,
as measured by Verstraete et al.~(1990).  This gives
\be
\label{eq:y0gra}
y_0(\Theta) = \frac{9 \times 10^{-3} \left(\Theta /W \right)^5}
{1+3.7 \times 10^{-2} \left(\Theta /W \right)^5}~~~,
\ee
with $W = 4.4 \eV$ (see Figure \ref{fig:yield_coronene}).  Draine
(1978) shows, in his Figure 1, that the bulk graphite yield measured by 
Feuerbacher \& Fitton (1972) is unusually low, and lies more than an 
order of magnitude below the yield for anthracene (Fujihira, Hirooka, 
\& Inokuchi 1973).
It is important to note that our equation (\ref{eq:y0gra}) gives yields
that substantially 
exceed those measured for bulk graphite.\footnote{For example, at
$h\nu = 10 \eV$, where Feuerbacher \& Fitton (1972) measured a yield of
$8 \times 10^{-3}$ for bulk graphite, our 
equation (\ref{eq:y0gra}) gives $y_0 = 2.7 \times 10^{-2}$, a factor 3.4
larger than the laboratory value.  For comparison,
the Bakes \& Tielens (1994) expression gives
$y_0 = 7.8 \times 10^{-2}$, an order of magnitude larger than the 
experimental value.}  
This apparent inconsistency could imply one or more 
of the following:  1.  The yield for bulk graphite differs
substantially from that for bulk coronene.  2.  The measured yields for
bulk graphite are wrong.  3.  Our prescription for 
$y_1(a, h\nu)$ is wrong.  Clearly, the adopted yields for carbonaceous grains
are highly uncertain.  

The situation for silicates is no better.  Feuerbacher et al.\ (1972) 
measured the yield for a powdered sample of lunar dust, and found that $Y$
decreases rapidly as photon energy is decreased below $14 \eV$.  However,
they noted that the use of the powdered sample might result in yields that 
are too low.  Thus, we choose $y_0$ for silicates such that the bulk yield 
somewhat exceeds the results of 
Feuerbacher et al.\ at $14 \eV$, but does not drop as rapidly for lower
values of $h\nu$:
\be
y_0(\Theta) = \frac{0.5 \left( \Theta /W \right)}{1+5 \left( \Theta /W
\right)}~~~,
\ee
with $W = 8 \eV$.

\begin{figure}
\epsscale{1.00}
\plotone{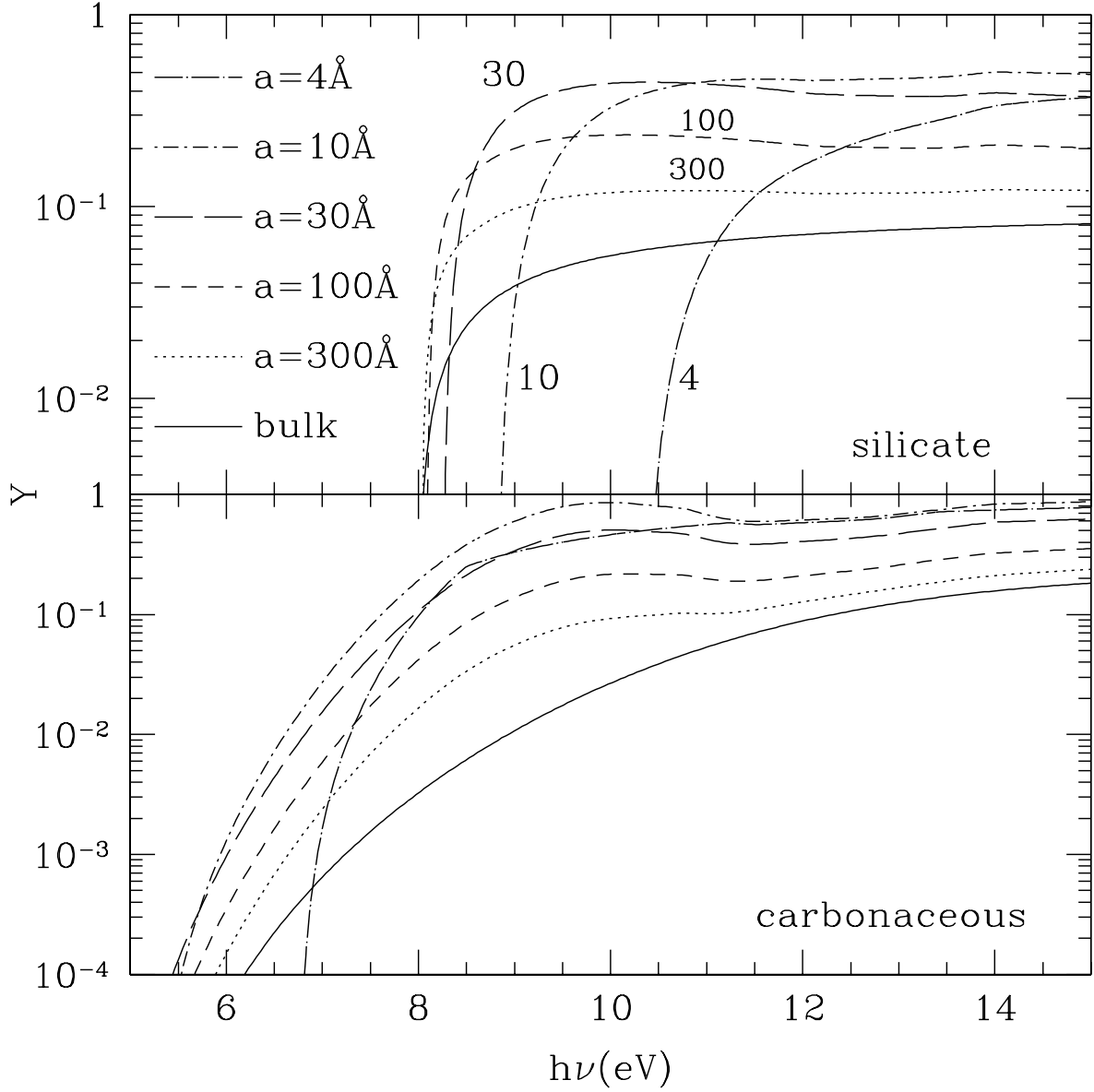}
\caption{
\label{fig:yield}
	Photoelectric yield $Y$ for 
neutral graphite and silicate grains as a 
function of incident photon energy $h\nu$, for several values of the grain
size $a$, as indicated.  
        }
\end{figure}

In Figure \ref{fig:yield} we plot $Y(h\nu)$ for neutral 
carbonaceous and silicate grains
of several sizes.

\subsubsection{Photoelectric Ejection of Attached Electrons (Photodetachment)}

When $Z<0$, the $-Z$ attached electrons occupy energy levels above the valence
band, if the latter is full in the neutral.  We take 
the photodetachment threshold energy $h\nu_\pdt$ to be
\be
h\nu_\pdt(Z<0) = EA(Z+1, a) + E_{\min}(Z,a) 
\ee
with electron affinities $EA$ 
given by equations (\ref{eq:EA}) and (\ref{eq:EAsil}), 
and $E_{\min}$ by (\ref{eq:Emin}).
Bakes \& Tielens (1994) neglected $E_{\min}(Z,a)$ and 
took $h\nu_\pdt = EA(Z+1, a)$.  We assume that $E=h\nu - h\nu_\pdt$ for 
photodetached electrons, since attached electrons lie in a narrow range of
energies.

We assume that an oscillator strength $f_\pdt$ is associated with
photodetachment transitions to the continuum.
The photodetachment cross section $\sigma_\pdt (h\nu)$ has been
measured for C$_6$F$_6^-$ (Christophorou, Datskos, \& Faidas 1994).
The measured cross section has considerable structure, but can be
roughly approximated by 
\be
\sigma_\pdt (Z,a) = -Z \frac{2\pi e^2 h f_\pdt}{3 m_e c \Delta E}
\frac{x}{(1+x^2/3)^2} ~~,
\ee
where $x \equiv (h\nu - h\nu_\pdt)/\Delta E$
and the peak in $\sigma_\pdt$ occurs at
$h\nu = h\nu_\pdt + \Delta E$.  We take $\Delta E = 3 \eV$ and 
oscillator strength $f_\pdt=0.5$; thus
\be
\sigma_\pdt (h\nu, Z, a) = 1.2\times10^{-17}\cm^2
	|Z| \frac{x}{(1+x^2/3)^2}~~~
{\rm for}~Z<0.
\ee

\subsection{Grain Charging}
    
Since the photoemission depends on the grain charge, it is necessary 
to know the distribution of charge states for the grains.  
In statistical equilibrium,
\be
f_Z(Z) [\Jpe(Z) +\Jion(Z)] = f_Z(Z+1) J_{e}(Z+1)~~~,
\label{balance}
\ee
where $f_Z(Z)$ is the probability for the grain charge to be $Ze$, $\Jpe$ is
the photoemission rate, $\Jion$ is the positive ion
accretion rate, and $J_{e}$ 
is the electron accretion rate.  

Bakes \& Tielens (1994) discuss the 
most positive and most negative charges that a grain could possibly acquire
($Z_{\max}e$ and $Z_{\min}e$, respectively).  
The most positive charge is one
proton charge more than the highest charge for which an electron can
be ejected, i.e. for which $h\nu_\pet = IP < h \nu_{\rm max}$, the maximum
photon energy in the radiation field ($=13.6 \eV$, in an H~I region).  Thus,
\be
Z_{\max}
 = {\rm{int}} \, \left[ \left( \frac{h\nu_{\max} - W}{14.4 \eV} \frac{a}
{\Angstrom} + \frac{1}{2} - \frac{0.3 \Angstrom}{a} \right) \left( 1 + 
\frac{0.3 \Angstrom}{a} \right)^{-1} \right]~~~,
\ee
where ${\rm{int}} \, [x]$ denotes the greatest integer less than $x$.
The minimum allowed charge $Z_{\min} e$ is the
most negative charge for which autoionization does not occur.  Bakes \& 
Tielens (1994) assume that this requires $EA(Z_{EA} + 1) > 0$.  However,
large grains can be charged more negatively without emitting a substantial
electron current when the tunneling probability is low.  We take the 
autoionization threshold potential $U_{\rm ait}$
to be that at which the electron current
is $\approx 10^{-6} {\rm s}^{-1}$; we evaluate the tunneling probability using 
the WKB approximation and assume an attempt frequency of 
$\approx 2 \times 10^8 {\rm s}^{-1} (\cm / a)$.  We find
\be
\frac{-U_{\rm ait}}{\rm V} \approx \cases{3.9 + 0.12 \left( a/
\Angstrom \right) + 2 \left( \Angstrom /a \right)
&, for carbonaceous\cr 2.5 + 0.07 \left( a/ \Angstrom
\right) + 8 \left( \Angstrom /a \right) &, for silicate\cr}~~~;
\label{eq:U_ait}
\ee
this agrees very well with Draine \& Salpeter's (1979) simple estimate of 
the potential at which field emission becomes important.  
The most negative allowed charge is then given by
\be
Z_{\min} = {\rm int} \, \left[ \frac{U_{\rm ait}}{14.4 {\rm V}} 
\frac{a}{\Angstrom} \right] + 1~~~.
\label{eq:Z_min}
\ee
For carbonaceous grains, equations (\ref{eq:U_ait})--(\ref{eq:Z_min})
give 
$Z_{\min}=0$ for $a<2.92\Angstrom$, 
$Z_{\min}=-1$ for $2.92\leq a < 5.75\Angstrom$,
$Z_{\min}=-2$ for $5.75\leq a < 8.40\Angstrom$.
For silicate grains,
$Z_{\min}=0$ for $a<2.40\Angstrom$,
$Z_{\min}=-1$ for $2.40\leq a < 6.96\Angstrom$,
$Z_{\min}=-2$ for $6.96\leq a < 10.8\Angstrom$.
By iteratively applying equation (\ref{balance}) 
and normalizing, $f_Z$ can be found for all
$Z$.  

The photoemission rate is given by
\be
\label{eq:jpe}
\Jpe = \pi a^2 \int_{\nu_\pet}^{\nu_{\max}} d\nu Y \Qabs(\nu) \frac{c u_{\nu}}
{h\nu}  +
\int_{\nu_\pdt}^{\nu_{\max}} d\nu \sigma_\pdt(\nu) \frac{c u_\nu}{h\nu}~~~,
\ee
where 
$\Qabs$ is the absorption efficiency,
$u_{\nu}$ is the radiation energy density per frequency interval, 
and $c$ is the speed of light.  The second term in equation (\ref{eq:jpe})
is only present when $Z<0$.  Bakes \& Tielens (1994) assumed
$\Qabs \propto a$, valid for grains with $a \ltsim 100 \Angstrom$.
Since we consider larger grains, we cannot adopt this approximation.
We evaluate $\Qabs$ for graphitic grains using the prescription of Li \& 
Draine (2000)\footnote{%
	Li \& Draine (2000) give optical properties for 
	neutral and ionized grains; we use the former when $Z=0$ and the latter
	when $Z \neq 0$.  The distinction is not critical, as
	in the vacuum ultraviolet 
	there is little difference between the ionized and neutral grain 
	absorption cross sections.
	} 
and for silicate grains
using a Mie theory code derived from BHMIE 
(Bohren \& Huffman 1983) with dielectric functions given by 
Draine \& Lee (1984) and Laor \& Draine (1993), but modified in the 
ultraviolet (Weingartner \& Draine 2000).    

The accretion rates are given by
\be
J_{i}(Z) = n_{i} s_i(Z) {\left( \frac{8kT}{\pi m_{i}}\right)}^{1/2} \pi a^2 
\tilde{J}(\tau_i,\xi_i)~~~,
\label{eq:ji}
\ee
where $n_i$ is the number density of species $i$, 
the sticking coefficient $s_i(Z)$ is the probability that species $i$
will transfer its charge if it reaches the surface of a grain of charge
$Ze$, $m_i$ is the particle mass,
$T$ is the gas temperature, and $\tilde{J}$ is a function of 
$\tau_i \equiv akT/{q_i}^2$ and 
$\xi_i \equiv Ze/q_i$ ($q_i$ is the charge of 
species $i$ and $k$ is the Boltzmann constant).   
Expressions for $\tilde{J}$ can be found in Draine \& Sutin (1987).
The electron and ion sticking coefficients $s_e$ and $s_i$ are discussed
below in \S \ref{sec:sticking}.

\section {Sticking Coefficients\label{sec:sticking}}

\subsection{Electrons}

A low-energy electron impinging on a macroscopic solid surface has some 
probability $p_{es}$ of elastic scattering.
We will assume that $p_{es}\approx 0.5$, so that the maximum possible value
of the sticking coefficient $s_e$ would be $\left(1-p_{es}\right)\approx 0.5$.

\subsubsection{Attachment to Neutral Grains}

We first consider electron attachment to neutral grains.  If the electron
affinity $EA > 0$ [as it is for $a> 3.5 \Angstrom$, which 
Weingartner \& Draine (2000) take  
to be the lower cutoff for the grain size 
distribution], then the approaching electron accelerates due to its 
polarization of the grain, arriving at the surface with a kinetic energy of
order $EA$.
Even if it enters the grain material [with probability $(1-p_{es})$],
the electron may fail to undergo an inelastic
scattering event, in which case it passes through the grain and 
returns to infinity.
The probability of undergoing inelastic scattering is approximately
$\left(1-e^{-a/l_e}\right)$, where $l_e\approx10\Angstrom$, the ``electron
escape length'' (discussed above), is roughly the mean free path
against inelastic scattering within the grain material.

If the impinging electron does undergo inelastic scattering, then it 
transfers some of its energy to internal degrees of freedom of the
grain; if the energy so transferred exceeds
the initial kinetic energy ($\sim 2kT$), then the electron is
trapped by the grain potential.  However, until the grain radiates
this energy away,
there is a nonzero probability per unit time that the internal degrees of
freedom of the grain will transfer enough energy back to the electron to
eject it.
Let $p_{\rm rad}$ be the probability of ``radiative stabilization'' of the
negatively charged grain, i.e. the probability of
radiating away $\sim kT$ of energy before the
electron is ejected.
We expect $p_{\rm rad}\rightarrow 1$ for macroscopic grains, 
but $p_{\rm rad} < 1$
for grains with a small number of internal degrees of freedom.
The electron attachment sticking coefficient 
can then be written as the product of three factors:
\be
s_e(Z=0)\approx \left(1-p_{es}\right)\left(1-e^{-a/l_e}\right) p_{\rm rad} ~~.
\ee
The stabilization probability $p_{\rm rad}$ can be estimated from the
electron affinity $EA$, the density of states of the excited negatively
charged grain created by the electron attachment, and
the probability per time of photon emission from
the excited grain (Allamandola, Tielens, \& Barker 1989; Tielens 1993).
However, since these parameters are poorly known, we will instead adopt
an empirical approach.

In Figure \ref{fig:s_eattach} we show sticking coefficients
$s_e = \langle \sigma v\rangle/[(8kT/\pi m_e)^{1/2}\pi a^2 \tilde{J}]$
for various hydrocarbon molecules, where $\tilde{J}(akT/e^2,0)$ 
is given by Draine \& Sutin (1987) and the
rate coefficients $\langle\sigma v\rangle$ for electron attachment are
determined experimentally.
Also shown in Figure \ref{fig:s_eattach} is a semiempirical fit: 
\be
\label{eq:s_e_0}
s_e(Z=0) = 0.5 \left(1-e^{-a/l_e}\right)
\frac{1}{1 + e^{(20 - N_{\rm C})}}  ~~~,
\label{eq:s_eattach}
\ee
where $N_{\rm C}$ is the number of atoms other than H in the molecule,
and $l_e=10^{-7}\cm$.

The measured sticking coefficients for $N_{\rm C} \le 20$ are in reasonable
agreement with our semiempirical fit (eq.~\ref{eq:s_eattach}).
It is notable that Tobita et al.\ (1992) found that in many cases the
sticking coefficient {\it increases} when the electron energy increases.
For example, the sticking coefficients for tetracene C$_{18}$H$_{12}$ and
perylene C$_{20}$H$_{12}$ would be larger by an order of magnitude for
gas temperature $T\approx 2000$K, bringing them up to or even above the 
values given by equation (\ref{eq:s_eattach}).

The measured $s_e$ for C$_{60}$ and C$_{70}$ at $T\le 500$K are remarkably
low, probably due to the
unusual rigidity and symmetry of these molecules.  For C$_{60}$, the
symmetry forbids capture of $s-$wave electrons, and the
centrifugal barrier to $p-$wave electrons results
in an apparent activation energy of 0.26eV (Tossati \& Manini 1994).
At high temperatures the sticking coefficients for $C_{60}$ and $C_{70}$
are close to the values given by equation (\ref{eq:s_eattach}).

We adopt equation (\ref{eq:s_eattach}) for $s_e(Z=0)$ for silicates
as well as for carbonaceous grains, since no laboratory data are available 
for silicates.  In this case, $N_{\rm C}$ is just a parameter defined by 
equation (\ref{eq:nc-a}).

\subsubsection{Attachment to Negatively Charged Grains}

As a grain acquires more electrons, the electron affinity decreases.  
We take the sticking coefficient to be zero when $Z=Z_{\min}$, since the
more negatively-charged state, even if it formed, would autoionize.
Thus,
\be
s_e(Z<0) = \cases{s_e(Z=0) &, $Z > Z_{\min}$\cr
0 &, $Z\leq Z_{\min}$\cr}~~~.
\ee

\subsubsection{Recombination with Positively Charged Grains}

Electrons arriving at positively charged grains are expected to recombine
provided that (1) they do not reflect elastically from the surface and (2)
they are able to scatter inelastically before traversing the
grain.  For small molecules the energy released generally results in
dissociation, but for
hydrocarbons this most likely results in only the loss of an H atom, with
the carbon skeleton remaining intact.
The sticking coefficient is therefore expected to be
\be
s_e(Z>0) \approx \left(1-p_{es}\right)
\left(1 - e^{-a/l_e}\right)
\approx 0.5\left(1 - e^{-a/l_e}\right)~~~.
\label{eq:s_erecomb}
\ee
Figure \ref{fig:s_erecomb} shows measured $s_e$ for
electron recombination with positive hydrocarbon 
ions, together with fitting function (\ref{eq:s_erecomb}).
The measured sticking coefficients generally 
agree with equation (\ref{eq:s_erecomb}) to
within about a factor of 2 except for the value measured for 
naphthalene C$_{10}$H$_8^+$, which is smaller than predicted by about
a factor of 4.  Why this should be smaller is unclear.

\subsection{Ions}

Ions of interest (H$^+$ and C$^+$) have large ionization potentials,
so we assume that they have a high probability of seizing an electron
if they arrive at the surface of a grain, whether charged or neutral.
While this assumption would be invalid for grains which are already highly
positively charged, in practice the rate of arrival of positive ions at
the surface of such a grain is in any case negligible.  We compute the
contribution of ions to the charging rate using
equation (\ref{eq:ji}) with $s_i=1$.

\section{Ambient Conditions\label{sec:ambient}}

\subsection{Radiation Fields\label{subsec:radfields}}

For some calculations, 
we adopt a blackbody spectrum for the radiation field, with
color temperature $T_{\rm c}$ and dilution factor $w$, so that $u_\nu =
4 \pi w B_\nu (T_{\rm c}) /c$.  It is convenient to characterize the
radiation intensity by $G \equiv
\urad^{\rm uv} / \uHab$, where $\urad^{\rm uv}$ is the 
energy density in the radiation field between $6 \eV$ and $13.6 \eV$
and $\uHab = 5.33 \times 10^{-14} \erg {\cm}^{-3}$ is the 
Habing (1968) estimate of the starlight energy density in this 
range.\footnote{
	For comparison, the interstellar radiation field estimated by
	Draine (1978) has $u=8.93\times10^{-14}\erg\cm^{-3}$ between
	6 and 13.6 eV, or $G=1.68$.
	}
The radiation is cut off at $13.6 \eV$.

For the diffuse ISM, we adopt
the average interstellar radiation field (ISRF) spectrum in
the solar neighborhood, as estimated by
Mezger, Mathis, \& Panagia (1982) and  Mathis, Mezger, \& Panagia (1983):
\be
\nu u_\nu^{\rm ISRF} = 
\cases{0 &, $h \nu > 13.6 \eV$\cr
3.328 \times 10^{-9} \E \left(h\nu/\eV\right)^{-4.4172} &,
$11.2 \eV < h \nu < 13.6 \eV$\cr
8.463\times10^{-13} \E \left(h\nu/\eV\right)^{-1} &,
$9.26 \eV < h \nu < 11.2 \eV$\cr
2.055\times10^{-14} \E \left(h\nu/\eV\right)^{0.6678} &,
$5.04 \eV < h \nu < 9.26 \eV$\cr
\left(4 \pi \nu/c\right) \sum_{i=1}^3 w_i B_\nu (T_i) &, $h \nu < 5.04 \eV$\cr}
~~~;
\label{eq:isrf}
\ee 
the dilution factors $w_i$ and blackbody temperatures $T_i$
are given in Table \ref{tab:ISRF}.
The total energy density in the ISRF of equation (\ref{eq:isrf}) is
$u=8.64\times10^{-13}\erg\cm^{-3}$, with
$u_{\rm rad}^{\rm uv}=6.07\times10^{-14}\erg\cm^{-3}$ in the
6-13.6 eV interval, or $G=1.13$.

The spectrum-averaged absorption efficiency factor is
\be
\langle \Qabs \rangle \equiv 
\frac{\int\Qabs u_\nu d\nu}{\int u_\nu d\nu}~~~,
\ee
where $\Qabs\pi a^2$ is the absorption cross section.
In Figure \ref{fig:qabs}
we display $\langle\Qabs\rangle$ for 
the ISRF and blackbody spectra with various values of $T_c$.  

\subsection{Scaling Law}

The photoelectric emission 
is dependent on the ambient conditions, which can be 
characterized by the shape of the radiation spectrum, the gas temperature 
$T$, and one additional parameter, depending on the ratio $G/n_e$,
which we take to be $G\sqrt{T}/n_e$.

Unless otherwise noted, we will 
display results for a blackbody spectrum with $\Tc = 3 \times 10^4 \K$,
cut off at $13.6 \eV$.  In Figure \ref{fig:chargedists}, charge distributions
are plotted for carbonaceous grains with various sizes in various 
environments.  In Figure \ref{fig:potential}, 
the average electrostatic potential, $\left<U\right>$,
is plotted as a function of grain size for two values of $T$, $100 \K$ and 
$1000 \K$, and five values of $G \sqrt{T} / n_e$, ranging from $10^2$ to 
$10^6 {\K}^{1/2}{\cm}^3$.  
We also provide results for conditions appropriate for the cold neutral
medium (ISRF, $T=100\K$, $n_e=0.03 {\cm}^{-3}$, $G\sqrt{T}/n_e = 
380\cm^3\K^{1/2}$)
and the warm neutral medium (ISRF, $T=6000\K$, $n_e=0.03 {\cm}^{-3}$, 
$G\sqrt{T}/n_e= 2900\cm^3\K^{1/2}$).
Note that for given $G\sqrt{T}/n_e$, the computed potentials show hardly
any dependence on $T$.  

For low values of $G \sqrt{T} / n_e$, the grains can be negatively charged.
This is especially so for grains with $a \ltsim l_e \approx 10 \Angstrom$, 
for which the photoemission rate scales approximately with $a^3$ rather than
$a^2$.  However, once $a \ltsim 4 \Angstrom$, the potential rapidly rises
to positive values; this is a consequence of the rapid decrease in $s_e$  
for grains with $Z \le 0$ as $a$ decreases in this range (eq. 
\ref{eq:s_e_0}).  The wiggles in the curves for high $G \sqrt{T} / n_e$ are 
due to saturation at the positive threshold charge $Z_{IP}$.  As the grain 
size increases, $f_Z(Z_{IP}) \rightarrow 1$, 
until the size at which $Z_{IP}$ increases by 1 is reached.  Thus, although
$\left< Z \right>$ increases monotonically with $a$, it plateaus until 
$Z_{IP}$ is incremented, so that $\left<U\right>$ decreases.  Once $Z_{IP}$
does increment, $\left<U\right>$ does not abruptly increase, because
the photoemission rate depends on $h\nu - h\nu_\pet$ and $h\nu - h\nu_\pdt$;
thus charge is gradually concentrated in the next higher charge state as 
$a$ increases.

To see why $G \sqrt{T} / n_e$ is a good parameter to 
describe grain charging, we make the approximation that the charge state on
any grain remains constant, i.e.
\be
J_e = \Jpe + \Jion~~~.
\ee
We neglect $\Jion$ by virtue of the large proton mass and consider a 
positively charged grain with $U \sim 1 \, {\rm V}$.  The Draine \& 
Sutin (1987) expression for $\tilde{J}$ for a grain with static
dielectric constant $\epsilon (0) \gg 1$ and $\xi_i < 0$ is
\be
\tilde{J} \approx \left( 1 - \frac{\xi_i}{\tau_i}\right) \left[ 1 +
{\left(\frac{2}{\tau_i - 2 \xi_i}\right)}^{1/2} \right]~~~,
\ee 
which expands, for $eU/kT \gg 1$, to give
\be
\tilde{J} \approx \frac{eU}{kT} \left(1+Z^{-1/2}\right)~~~.
\label{expanded}
\ee
We express
\be
\Jpe \propto G \left<\Qabs\right> a^2 g(U, \, a)~~~,
\ee
where $\left<\Qabs\right>$ 
is the average value of $\Qabs$ over the range of absorbed
photon energies.  Here $g(U, \, a)$ must be a decreasing function of $U$  
and $g$ depends on $a$ through the size dependence of the yield $Y$
(see equations \ref{yeqn} and \ref{fyeqn}).  Thus, 
\be
\frac{U\left(1+Z^{-1/2}\right)}{g(U, \, a)} \propto \left<\Qabs\right> 
\frac{G \sqrt{T}}{n_e}~~~.
\label{eqn:scaling}
\ee

Note that this result applies only when $G / n_e$ is large enough
(for given $T$ and $a$)
to keep the grains strongly positively charged (with $eU\gtsim kT$).  For
low values of $G / n_e$, $\Jion$ plays a significant role in the
grain charging.  This introduces an additional dependence on the ionization
fraction $x \equiv n_e / \nH$, since for the lowest $x$ (i.e.~$x \lesssim
2 \times 10^{-4}$), the ions are predominantly C$^+$, whereas H$^+$ dominates
for higher $x$, and $\Jion$ depends on the ion mass.  We assume  
H$^+$ in our calculations for the CNM and WNM and C$^+$ in our other
calculations.  For the values of $G \sqrt{T} / n_e$ and $T$
considered here, the
dependence on ion mix is only significant for the smallest grains, and 
even then extreme changes in the ion mix lead to only modest changes 
in the results.  In Figure \ref{fig:critpar} we display 
$(G \sqrt{T} / n_e)_0$, the value of the charging parameter for which 
$\langle Z \rangle = 0$, for carbonaceous and silicate grains, a blackbody 
radiation field with $\Tc = 3 \times 10^4 \K$, and various gas
temperatures.  As $a$ decreases beyond $\approx 4 \Angstrom$ it becomes very
difficult for the grains to charge negatively, because $s_e$ becomes very
small for $Z \le 0$; thus the curves plunge sharply there.  

Incidentally, the presence of $\left<\Qabs\right>$ in equation 
(\ref{eqn:scaling}) explains the maxima in $\left<U\right>$ at
$a \sim 100 \Angstrom$ (see Figure \ref{fig:potential}), 
since $\Qabs$ peaks for $2 \pi a \sim \lambda$, and
the photoelectric emission is primarily from radiation with 
$\lambda \sim 1000 \Angstrom$.  

\section{Photoelectric Heating Efficiencies\label{sec:peheateff}}

The gas heating rate per grain due to photoelectric emission is given by
\be
\Gampe^\prime (a) = \sum_Z f_Z(Z) \left[ \Gamma_{\rm pe, v}^{\prime}
(a) + \Gamma_{\rm pd}^{\prime}(a) \right]~~~.
\ee
The contribution from the photoemission of valence electrons is
\be
\Gamma_{\rm pe, v}^{\prime}(a) = \pi a^2
\int_{\nu_\pet}^{\nu_{\rm max}} d\nu Y \Qabs
\frac{c u_{\nu}}{h\nu} \int_{E_{\rm min}}^{E_{\rm max}} dE f_E(E) E~~~,
\ee
where $E_{\rm min} = 0$ when $Z \ge 0$ and is given by equation (\ref{eq:Emin})
when $Z < 0$, and $E_{\rm max} = h\nu - h\nu_\pet + E_{\rm min}$.
The photoelectron energy distribution $f_E(E) = f_E^0(E)/y_2$ (see eq.
\ref{eq:parab} and the discussion following it).  
When $Z<0$, the contribution from photodetachment is given by
\be
\Gamma_{\rm pd}^{\prime}(a) = \int_{\nu_\pdt}^{\nu_{\rm max}} d\nu  
\sigma_{\rm pdt}(\nu) \frac{c u_{\nu}}{h\nu} 
(h\nu - h\nu_\pdt + E_{\rm min})~~~.
\ee

We compute the total efficiency 
for conversion of absorbed radiation into gas heating,
\be
\epsilon_\Gamma(a) = \frac{\Gampe^\prime (a) -\Lamgr^\prime (a)}
{\pi a^2 c u_{\rm rad} \langle Q_{\rm{abs}} \rangle} ~~~.
\ee
Here $\Lamgr^\prime (a)$ 
is the rate of energy removal from the gas due to the accretion of 
charged particles onto the grain, and is given by
\be
\Lamgr^\prime (a)
=\sum_{i}n_{i} s_{i} {\left( \frac{8kT}{\pi m_{i}}\right)}^{1/2} \pi a^2 
\tilde{\Lambda}(\tau_i, \xi_i) k T~~~,
\ee
where the sum runs over electrons and ions and  $\tau_i$ and $\xi_i$ are 
defined below equation (\ref{eq:ji}).
Expressions for $\tilde{\Lambda}$ can be found in Draine \& Sutin (1987).

Figures \ref{fig:gammapegra} through \ref{fig:gammapetc} show 
results for the gas heating efficiency for the 
same conditions for which potentials were displayed 
in Figure \ref{fig:potential}.  
The heating efficiency generally decreases as grain size $a$ increases.  For
low values of $G \sqrt{T}/n_e$, this results primarily from the 
size-dependent yield factor $y_1$ (equation \ref{fyeqn});
the heating efficiency levels off at $a \sim 10^3 \Angstrom$, where $y_1$
levels off.  For higher $G \sqrt{T}/n_e$, high grain potentials quench
the photoemission and result in much lower heating efficiencies; in these
cases the smallest grains (with $a \ltsim 100 \Angstrom$), with somewhat 
lower
potentials (see Figure \ref{fig:potential}), dominate the photoelectric
heating. 
The heating efficiencies drop dramatically when $a \ltsim 4 \Angstrom$, as
$s_e(Z \le 0)$ decreases rapidly and the grains charge positively.

\section{Net Photoelectric Heating Rate\label{sec:netheat}}

In this section we integrate the photoelectric heating efficiency over 
grain size distributions to find net heating rates.  
Cardelli, Clayton, \& Mathis (1989) found that
the extinction curve 
-- and therefore the grain size distribution --
varies depending on the environment through which 
the starlight passes, and that the variation can be roughly parameterized 
by $R_V \equiv A_V/E_{B-V}$, the ratio of visual extinction to reddening.
For the diffuse ISM, $R_V \approx 3.1$; higher values are observed for 
dense clouds.

Emission in the 3 to 60$\micron$ range, presumably generated by grains
small enough to reach temperatures of 30 to 600$\K$ or more upon the 
absorption of a single starlight photon (see, e.g.~Draine \& Anderson 1985),
imply a population of very small grains (with $a < 50 \Angstrom$).  
The non-detection of
the $10 \micron$ silicate feature in emission from diffuse clouds
(Mattila et al.\ 1996; Onaka et al.\ 1996) appears to rule out silicate grains
as a major component of the $a \lesssim 15 \Angstrom$ population (Li \&
Draine 2000).  Emission features at 3.3, 6.2, 7.7, 8.6,
and $11.3 \micron$ (see Sellgren 1994 for a review) have been identified 
as C-H and C-C stretching and bending modes in polycyclic aromatic 
hydrocarbons (L\'{e}ger \& Puget 1984), suggesting that the carbonaceous 
grain population extends down into the molecular regime.  
The C abundance\footnote{By ``abundance'', we mean the number of atoms of 
an element per interstellar H nucleus.} 
$b_{\rm C}$ in the very small grain population is still 
uncertain; comparison of the observed diffuse galactic infrared and 
microwave emission with detailed model calculations for grains heated by 
galactic starlight imply $b_{\rm C} \approx 2$--$6 \times 10^{-5}$ 
(Li \& Draine 2000; Draine \& Lazarian 1998a,b; Draine \& Li 2000).

Weingartner \& Draine (2000, hereafter WD00)
obtained size distributions with various values of $b_{\rm C}$
which reproduce the average observed extinction for lines of sight 
characterized by $R_V=3.1$, 4.0, and 5.5.  Some of their distributions, 
designated ``case A'', were constructed so as to minimize the use of C and 
Si.  For $R_V=4.0$ and 5.5, ``case B'' distributions use the same amount of
C and Si as those with $R_V=3.1$.  Li \& Draine (2000) found that the diffuse
galactic infrared emission is best reproduced with the WD00
size distribution with $R_V=3.1$ and $b_{\rm C}=6 \times
10^{-5}$.  For this distribution, the $2175 \Angstrom$ hump in the extinction
curve is entirely due to the ultrasmall grain population.  If this is 
generally the case throughout the ISM, then we would have $b_{\rm C} =
4 \times 10^{-5}$ when $R_V=4.0$ and $b_{\rm C}=3\times 10^{-5}$ when
$R_V=5.5$.  We will usually display integrated heating rates computed for
the size distributions with these (largest allowed) values of $b_{\rm C}$.

In Figure \ref{fig:gr_dist_htg} we plot the net total gas heating per H 
nucleus and per Habing flux,
\be
\frac{\Gamma_{\rm tot}}{G \nH} = \sum_{c,s}
\int_{a_{\rm min}}^{a_{\rm max}} 
\frac{\Gamma_{\rm pe}^\prime - \Lamgr^\prime}{G} \frac{1}{\nH}
\frac{dn_{\rm gr}}{da} da 
\ee
as a function of $G \sqrt{T} / n_e$ for our favored distributions from 
WD00 with $T = 100 \K$ and $T_c = 3 \times 10^4 \K$.  The sum is over the 
carbonaceous and silicate populations.  At a single $G \sqrt{T}/n_e$,
the heating rates obtained for the
several distributions vary by a factor of $\approx \,$2--4, showing the 
sensitivity of $\Gamma_{\rm tot}$ to
the grain size distribution.  Of course, the highest heating 
rates are for the distribution with $R_V=3.1$ and $b_{\rm C}=6\times 10^{-5}$,
which contains the largest population of very small grains.
For comparison, we also plot the Bakes \& Tielens (1994) result.  
When $G \sqrt{T}/n_e \gtsim 10^4 \K^{1/2} \cm^3$, their curve lies
above our ($R_V=3.1$, $b_{\rm C}=6\times 10^{-5}$) curve, even though they
assumed a smaller population of very small grains; this is because they
adopted larger electron sticking coefficients than we do.
In Figure \ref{fig:gr_dist_htg_isrf} we 
show $\Gamma_{\rm tot}/G \nH$ for dust exposed to the ISRF; here we 
consider typical diffuse cloud dust with $R_V = 3.1$, for two possible values
of $b_{\rm C} = 0$ and $6\times 10^{-5}$,
and gas temperatures of $100$ and $6000 \K$.  Note that
the net heating when $T=100 \K$ does not differ greatly from that displayed
in Figure \ref{fig:gr_dist_htg} and that 
the grains have a net cooling effect for low values of $G \sqrt{T}/n_e$ 
when $T=6000 \K$.  

The photoelectric heating rate for the WD00
grain size distributions is fairly well reproduced by the following function:
\be
\Gamma_{\rm pe} \equiv \sum_{g,s}
\int_{a_{\rm min}}^{a_{\rm max}} \Gamma_{\rm pe}
^{\prime}(a) \frac{dn_{\rm gr}}{da} da
= 10^{-26} \frac{\rm{erg}}{{\rm s}}
G \nH \frac{C_0 + C_1 T^{C_4}}{1 + C_2 (G \sqrt{T}/n_e)^{C_5}
\left[ 1 + C_3 (G \sqrt{T}/n_e)^{C_6} \right]},
\label{eqn:totheatfit}
\ee
where $T$ is in $\K$ and $G \sqrt{T}/n_e$ is in $\K^{1/2} \cm^3$.
Values for the seven parameters in equation (\ref{eqn:totheatfit})
are given in Table \ref{tab:totheat}, along with the largest fractional 
error, err, for $10 \le T \le 10^4 \K$
and $10^2 \le G \sqrt{T}/n_e \le 10^6 \K^{1/2} \cm^3 $.
For most cases shown,
we consider a blackbody spectrum with $\Tc = 3 \times 10^4 \K$;
we also consider the ISRF for $R_V = 3.1$.  

Bakes \& Tielens (1994) neglect heating from grains with $a > 100
\Angstrom$.  We find that this is usually a good approximation; large grains 
contribute significantly only for 
low $G \sqrt{T}/n_e$ and size distributions with few small grains.
This is 
demonstrated in Table \ref{tab:totheat}, where we give the fraction 
$h_s$ of the total heating due to grains with $a < 100 \Angstrom$, for 
$G \sqrt{T}/n_e = 10^2 \K^{1/2} \cm^3$ and $T = 100 \K$.  
For example, 
for $R_V = 3.1$, $b_{\rm C} = 0.0$, and $\Tc = 3 \times 10^4 \K$, 
$25 \%$ of the heating is contributed by grains with $a > 100 \Angstrom$.
Bakes \& Tielens also neglect the heating contributed by silicate grains.
We find that silicates contribute $\ltsim 25 \%$ of the total heating,
with the maximum occuring for the distributions with the fewest
very small grains.

The rate of cooling due to charged particle collisions is significant, 
compared with the photoelectric heating rate, when $T \gtsim 10^3 \K$.  
The following approximation is fairly accurate 
when $10^3 \le T \le 10^4 \K$ and 
$10^2 \le G \sqrt{T}/n_e \le 10^6 \K^{1/2} \cm^3$:
\be
\Lamgr = 10^{-28} \, {\rm erg} \cm^3 \, {\rm s}^{-1} \,  
n_e \nH T^{(D_0+D_1/x)}
\exp \left( D_2 + D_3 x - D_4 x^2 \right)~~~,
\label{eqn:totcoolfit}
\ee
where $x \equiv \ln (G \sqrt{T}/n_e)$,
$T$ is in $\K$, and $G \sqrt{T}/n_e$ is in $\K^{1/2} \cm^3$.  In Table
\ref{tab:coolpars} we give the values of $D_i$ and the maximum fractional
error, err, for the above range of $T$ and $G \sqrt{T}/n_e$.

\section{Heating and Cooling in H~II Regions\label{sec:HIIregions}}

\subsection{Str\"{o}mgren Spheres}

Lyman continuum radiation from hot stars photoionizes the surrounding gas,
resulting in an H~II region.
Studies of the heating in H~II regions have found that photoelectric 
emission from dust can be important, compared with photoionization of H
(Maciel \& Pottasch 1982; Oliveira \& Maciel 1986; Maloney, Hollenbach, 
\& Townes 1992).  In these studies, it was  
assumed that the photoelectric yield $Y$ and
absorption efficiency factor $\Qabs$ are independent of grain size and 
photon energy.  Here, we apply our more detailed photoemission model.

In H~II regions, the radiation field includes photons with $h \nu > 
13.6 \eV$.  To see the possible importance of these photons, 
we calculate integrated photoelectric heating and recombination cooling 
rates for blackbody spectra with no upper cutoff energy, adopting 
$\Tc = 3.5$ and $4.5 \times 10^4 \K$.
Of course, at any given location in an H~II region, there will be a 
break in the spectrum at $13.6 \eV$, due in part to absorptions along the
path to the star and in part to the break in the stellar spectrum itself.
Thus, spectra with no break and with a cutoff at $13.6 \eV$ should
bracket the range applicable in H~II regions.  

We take $n_e / \nH = 1$ and $T = 9000 \K$ for the gas.
In order to estimate likely $G/\nH$ values, we consider
an O9$\,$V exciting star and a point at the half-mass radius (i.e. at the 
distance from the star for which half of the H~II region volume is 
enclosed).  The radius of the H~II region (the ``Str\"{o}mgren radius'') is
found by balancing ionizations against recombinations:
\be
R_S = \left( \frac{3}{4 \pi} \frac{\dot{N}_{\rm Ly}}{\alpha \nH^2}
\right)^{1/3},
\ee
where $\dot{N}_{\rm Ly}$ is the rate at which the star produces ionizing 
photons and the case B recombination coefficient 
$\alpha \approx 
2.6\times 10^{-13}T_4^{-0.8}\cm^3 \, {\rm s}^{-1}$, where
$T_4 \equiv T/10^4 \K$ (Osterbrock 1989).
Since $R_S \propto \nH^{-2/3}$, $G/\nH \propto \nH^{1/3}$.  We take
$\dot{N}_{\rm Ly} = 3.63 \times 10^{48} \, {\rm s}^{-1}$, luminosity
$L = 4.4 \times 10^{38} \erg \, {\rm s}^{-1}$, and effective temperature
$T_{\rm eff} = 3.59 \times 10^4 \K$ (Vacca, Garmany, \& Shull 1996).
For $\nH = 0.1$ ($10^3$) $\cm^{-3}$, $G/\nH \approx 0.5$ (10.)
$\cm^3$.

In Tables \ref{tab:heatHII} and \ref{tab:coolHII}, we give 
$\Gamma_{\rm pe}/G \nH$ and $\Lamgr/G \nH$ for $G/\nH = 0.1$, $1$, and
$10 \cm^3$, both with and without the cutoff at $13.6 \eV$, for several 
of the WD00 grain size distributions.  The heating rate typically
increases by a factor of a few when the cutoff is removed,
and the cooling rate is modestly affected, due to changes in the charging. 
When $G/\nH = 1.0\cm^3$,
grains have a net cooling (heating) effect for radiation with(out) a cutoff 
at $13.6 \eV$; grains always have a net heating effect when 
$G/\nH = 10.\cm^3$.  When $G/\nH < 0.1 \cm^3$ the grains
are negatively charged, so that $\Gamma_{\rm pe}/G \nH \approx 
{\rm constant}$, and $\Lamgr/G \nH \propto (G/\nH)^{-1}$, roughly.    

The net steady-state heating rate due to photoionization is, in the notation
of Spitzer (1978),
\be
\label{eq:gamma_pi}
\Gamma_{\rm pi} \approx \alpha n_e^2 
\left(\langle\psi\rangle k \Tc - (\chi_2/\phi_2) kT
\right)
\approx 1.5 \times 10^{-24}
n_e^2
T_4^{-0.8}
\left(\frac{\langle\psi\rangle\Tc-(\chi_2/\phi_2) T}{42000\K}\right)
\erg \cm^3 \s^{-1}.
\ee
where
$\langle\psi\rangle k\Tc$ is the mean kinetic energy per photoelectron,
and $(\chi_2/\phi_2) kT$ is the mean kinetic energy per recombining electron
for case B recombination;
$\langle\psi\rangle\approx 1.067 (\Tc/10^4\K)^{0.23}$ for
$16000\ltsim \Tc \ltsim 64000\K$,
and $\chi_2/\phi_2\approx0.67T_4^{-0.13}$ (Spitzer 1978).
Comparing with the entries in Tables \ref{tab:heatHII}
and \ref{tab:coolHII}, we find that the net
heating from dust can be comparable to or exceed the 
photoionization heating when $G/\nH \gtsim 5 \cm^3$.

In an actual H~II region, the grain size distribution could be a function
of distance from the exciting star, and could differ substantially from 
those found by WD00.  The very small grains might be destroyed in
the harsh ionizing environment and the large grains might drift away 
from the star.  A detailed study of the contribution of photoelectric 
heating in H~II regions would have to account for these effects.  

\subsection{The Warm Ionized Medium}

The warm ionized medium (WIM), also known as the diffuse ionized gas (DIG), 
extends to distances $|z| > 1 \, {\rm kpc}$ above the Galactic midplane 
(see Reynolds 1990, 1993 for reviews).  It is not yet clear how the WIM is
ionized and heated.  
The only known source of ionization with adequate power to maintain the WIM
is Lyman continuum radiation from O stars, but it is not clear how the
ionizing photons can reach locations with high $|z|$, since atomic gas is 
highly opaque to these photons.
Observed emission line intensity ratios can be approximately reproduced by
models in which photoionization by a dilute radiation field dominates 
(Domg\"{o}rgen \& Mathis 1994).  However, such models fail to account for
observed variations in line intensity ratios with $|z|$ (see Reynolds, 
Haffner, \& Tufte 1999 and references therein).

Haffner, Reynolds, \& Tufte (1999) found that the line intensity variations
could be due to variations in the gas temperature; specifically, they 
suggested an increase from $T \approx 7000 \K$ at $|z| \approx 500 \pc$
to $T \approx 10^4 \K$ at $|z| \approx 1500 \pc$.  Reynolds et al.~(1999) 
noted that the variations are more generally correlated with the electron 
density, suggesting a supplemental heat source that dominates photoionization
heating at low densities.  They found that a supplemental heat source 
$\Gamma \sim 10^{-25} \nH \erg \s^{-1}$ would result in a temperature profile
$T(|z|)$ which could account for the observed line intensity ratio variations,
and suggested photoelectric heating by dust as a possibility.

In Figure \ref{fig:t_wim}, we plot the $T(|z|)$ profile inferred by 
Reynolds et al.~(1999) from observed values of the [N II]/H$\ \alpha$ line
intensity ratio.  In this section we will investigate whether the combination
of heating by photoionization and photoemission from dust can reproduce this
profile.  Reynolds et al.~find that the electron density profile can be 
approximated by
\be
n_e (|z|) = 0.125 T_4^{0.45} f^{-0.5} \exp (-|z|/1 \, {\rm kpc}) \cm^{-3}~~~,
\ee
where $T_4 = T/10^4 \K$ and $f$ is the WIM volume filling fraction. 
Reynolds et al.\ considered two cases for $f(|z|)$:  $f(|z|)=0.2$ and 
$f(|z|)= 0.1 \exp (|z|/0.75 \, {\rm kpc})$ (Kulkarni \& Heiles 1987).  We
calculate temperature profiles for four different cases (A--D), 
as summarized in 
Table \ref{tab:wim}, employing the above two prescriptions for the filling 
factor and two grain size distributions, for $R_V=3.1$ and $b_{\rm C} = 0$
and $6 \times 10^{-5}$.  We take the 
net heating rate due to photoionization
$\Gamma_{\rm pi} = 1.5 \times 10^{-24} n_e^2 T_4^{-0.8} \erg \cm^3 \s^{-1}$ 
(appropriate for 
$\Tc\approx3.5 \times 10^4 \K$; equation \ref{eq:gamma_pi}) and 
cooling rate (excluding grain collisional cooling) 
$\Lambda_{\rm gas}= 3.0 \times10^{-24}T_4^{1.9} n_e^2 \erg \cm^3 \s^{-1}$, 
which approximates
the cooling function for ionized gas given in Figure 3.2 of Osterbrock 
(1989), and find $T$ satisfying
\be
\Gamma_{\rm pi} + \Gampe = \Lambda_{\rm gas} + \Lamgr ~~.
\ee

We plot the temperature profiles for the four cases A--D
in Figure \ref{fig:t_wim}; we have adjusted the values of $G$ so that 
the curves roughly lie on top of the Reynolds et al.~curve.  In Figure 
\ref{fig:wim2}, we plot the associated values of the charging parameter
$G \sqrt{T}/n_e$ and the net grain heating
$\Gamma_{\rm tot}/\nH$ as functions of $|z|$; note
that $\Gamma_{\rm tot}/\nH$ is not constant, since $T$ and $G \sqrt{T}/n_e$
vary with $|z|$.  For each of the cases A--D, the temperature profiles are 
too shallow compared with the Reynolds et al.~curve.  One might expect 
steeper curves to result if $G$ is increased, although this might require
reducing $\Gamma_{\rm pi} / n_e^2$ to unreasonably small values.  However,
this is not usually the case, because the photoelectric heating 
efficiency begins to 
drop rapidly when $G \sqrt{T} / n_e \gtsim 10^3 \K^{1/2} \cm^3$.  
We can obtain
a reasonable match to the Reynolds et al.~curve by altering case D
somewhat, so that 
$\Gamma_{\rm pi} = 0.7 \times 10^{-24} n_e^2 T_4^{-0.8}
\erg \cm^3 \s^{-1}$ and $G=0.27$ 
(case E).\footnote{%
	This could arise, for example, for $\Tc\approx26000\K$.
	}
Thus,
we can reproduce the Reynolds et al.~temperature profile by adopting 
a large
population of very small grains and the Kulkarni \& Heiles (1987) formula for
the WIM filling factor as a function of $|z|$.  
Other heating mechanisms, such as the dissipation of interstellar
turbulence (Minter \& Spangler  1997), could also contribute, but do not
appear to be required to account for the observed temperatures.

\section{Summary\label{sec:summary}}

We have presented a model for photoelectric emission from dust.  Here we 
first summarize how our model differs from the recent study by
Bakes \& Tielens (1994) and the uncertainties that remain.

We have re-evalulated the photoemission threshold energies for small grains,
based in part on an empirical approach employing laboratory data for 
ionization potentials and electron affinities of PAH molecules.  For 
negatively charged grains, we consider both photoemission from the valence 
band and the removal of attached electrons, and set the photoemission and 
photodetachment threshold energies equal to the minimum excitation energy for 
which an electron can effectively tunnel across the Coulomb barrier.  Bakes
\& Tielens take the emission threshold equal to the electron affinity,
which is an underestimate.  Motivated by the PAH hypothesis, Bakes \& 
Tielens adopt a disk geometry for the smallest grains.  The geometry affects
the ionization potential through the capacitance, but this effect is small.
We assume spherical grains for simplicity, and thereby retain consistency
with the calculation of the charged particle accretion rates $J_e$ and
$J_{\rm ion}$.  

We treat the variation of the photoelectric yield with grain charge
and the distribution of photoelectron energies in a consistent manner
(eq. \ref{eq:y1}).  Whereas Bakes \& Tielens adopt a delta function
for the distribution of photoelectron energies, we adopt a parabolic form
(eq. \ref{eq:parab}).  We adopt a yield function for carbonaceous grains
which, like that of Bakes \& Tielens, approximately reproduces the 
measured ionization yield of coronene for grains with $a \approx 4 \Angstrom$,
but which agrees better with experimental determinations of the bulk graphite
yield.  Still, our bulk yields substantially exceed the experimental results;
see the discussion following equation (\ref{eq:y0gra}).  In computing the
enhancement factor for the yield of small particles, we evaluate the 
photon attenuation length $l_a$ as a function of wavelength (eq.
\ref{eq:l_a}), rather than adopting $l_a = 100 \Angstrom$, independent of
wavelength, as Bakes \& Tielens do.  Recall, however, that the expression 
for the size-dependent yield enhancement (eq. \ref{fyeqn}) is based on
a simple theoretical model; laboratory investigation of photoemission from 
microscopic particles is still in its infancy.  

\begin{figure}
\epsscale{1.00}
\plotone{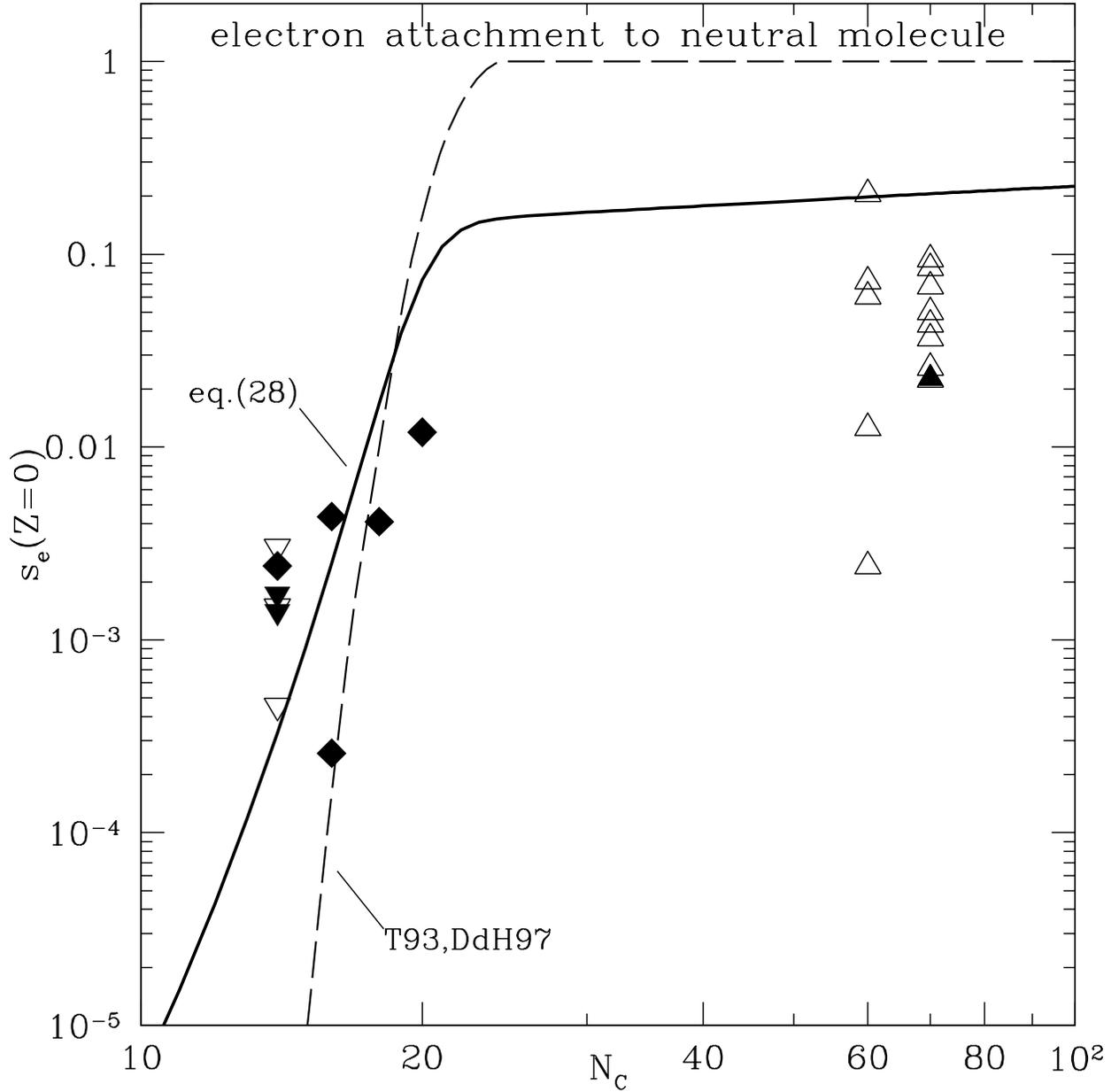}
\caption{
\label{fig:s_eattach}
	Sticking coefficients $s_e$ for electron
	attachment to neutral molecules, as a function of $N_{\rm C}$,
	the number of atoms other than H (mostly C).
	Diamonds: 
	acridine C$_{13}$H$_9$N,
	fluoranthene C$_{16}$H$_{10}$,
	pyrene C$_{16}$H$_{10}$,
	tetracene C$_{18}$H$_{12}$, and
	perylene C$_{20}$H$_{12}$
	(Tobita et al.\ 1992). 
	Triangles: 
	anthracene C$_{14}$H$_{10}$ (Tobita et al.\ 1992; 
	Canosa et al.\ 1994; Moustefaoui et al.\ 1998),
	C$_{60}$ (Smith et al.\ 1993), and
	C$_{70}$ (Spanel \& Smith 1994).
	Filled symbols are for $T=300$K.
	Open symbols for C$_{60}$ and C$_{70}$ are for $T=500 - 4500$K.
	Open symbols for anthracene are for $T=48 - 170$K
	(Moustefaoui et al.\ 1998).
	Solid curve: adopted empirical fit given by eq. (\ref{eq:s_eattach}).
	Broken curve: $s_e$ adopted by Tielens (1993) and 
	Dartois \& d'Hendecourt (1997).
	}
\end{figure}

\begin{figure}
\epsscale{1.00}
\plotone{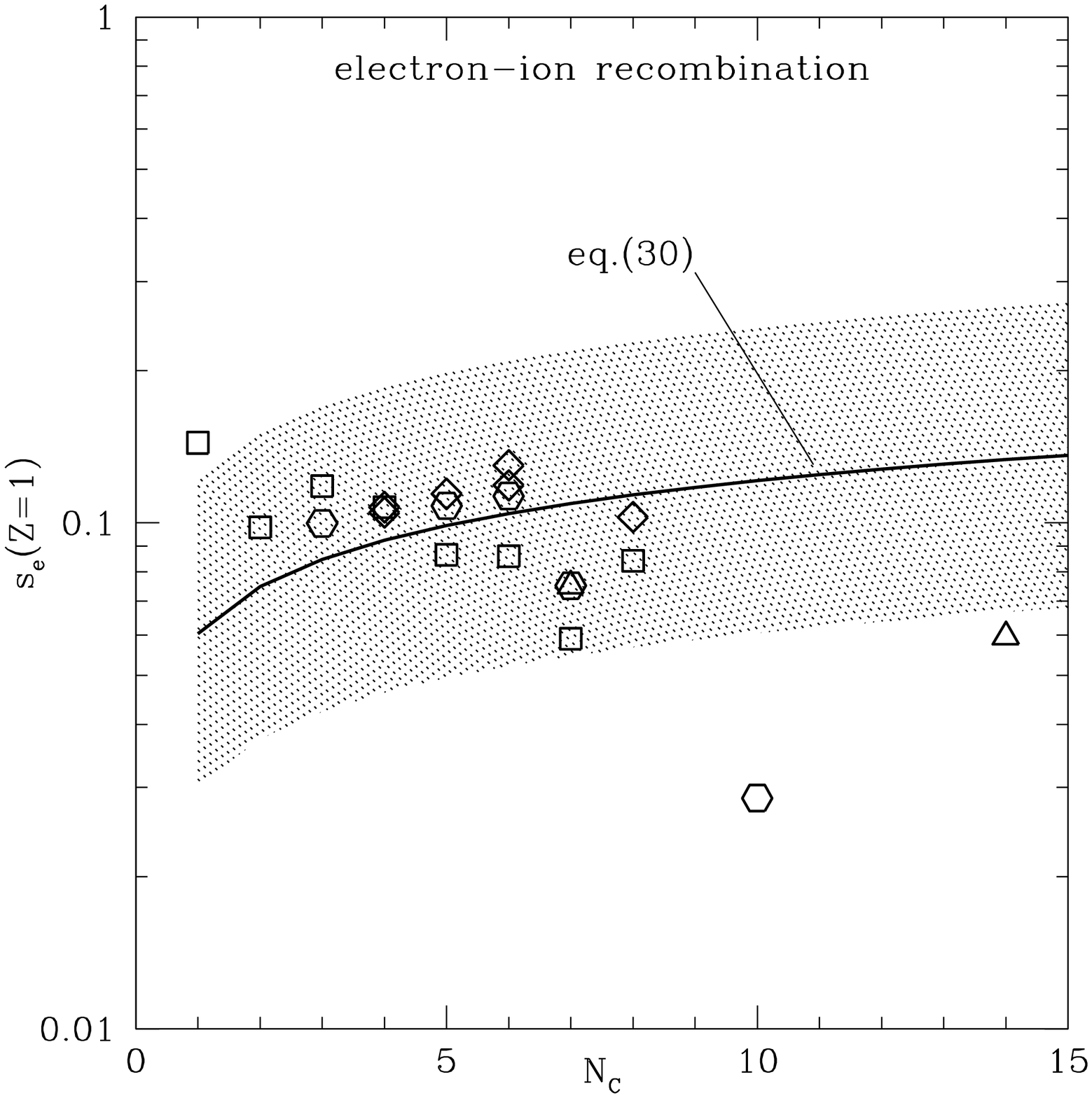}
\caption{
\label{fig:s_erecomb}
	Sticking coefficients $s_e$ for electron
	recombination with positive ions.
	Hexagons: $T=300$K experimental results of 
	Abouelaziz et al.\ (1993) for C$_3$H$_3^+$,
	C$_5$H$_3^+$,
	C$_6$H$_6^+$,
	C$_7$H$_5^+$, and
	C$_{10}$H$_8^+$. 
	Squares: $T=300$K results of Lehfaoui et al.\ (1997) for alkanes
	CH$_5^+$,
	C$_{2}$H$_{5}^+$,
	C$_{3}$H$_{7}^+$,
	C$_{4}$H$_{9}^+$,
	C$_{5}$H$_{11}^+$,
	C$_{6}$H$_{13}^+$,
	C$_{7}$H$_{15}^+$, and
	C$_{8}$H$_{17}^+$.
	Triangle: $T=300$K results of Rowe et al.\ (1995)
	for
	C$_7$H$_8^+$ and
	C$_{14}$H$_{10}^+$ phenanthrene.
	Diamonds: $T=300$K results of Rebrion-Rowe et al.\ (1998) for
	C$_{4}$H$_{5}^+$
	C$_{4}$H$_{11}^+$
	C$_{5}$H$_{9}^+$
	C$_{6}$H$_{4}^+$
	C$_{6}$H$_{5}^+$
	C$_{8}$H$_{7}^+$.
	Solid curve: adopted empirical fit given by eq. (\ref{eq:s_erecomb});
	shaded region is within factor of two.
	}
\end{figure}

\begin{figure}
\epsscale{1.00}
\plotone{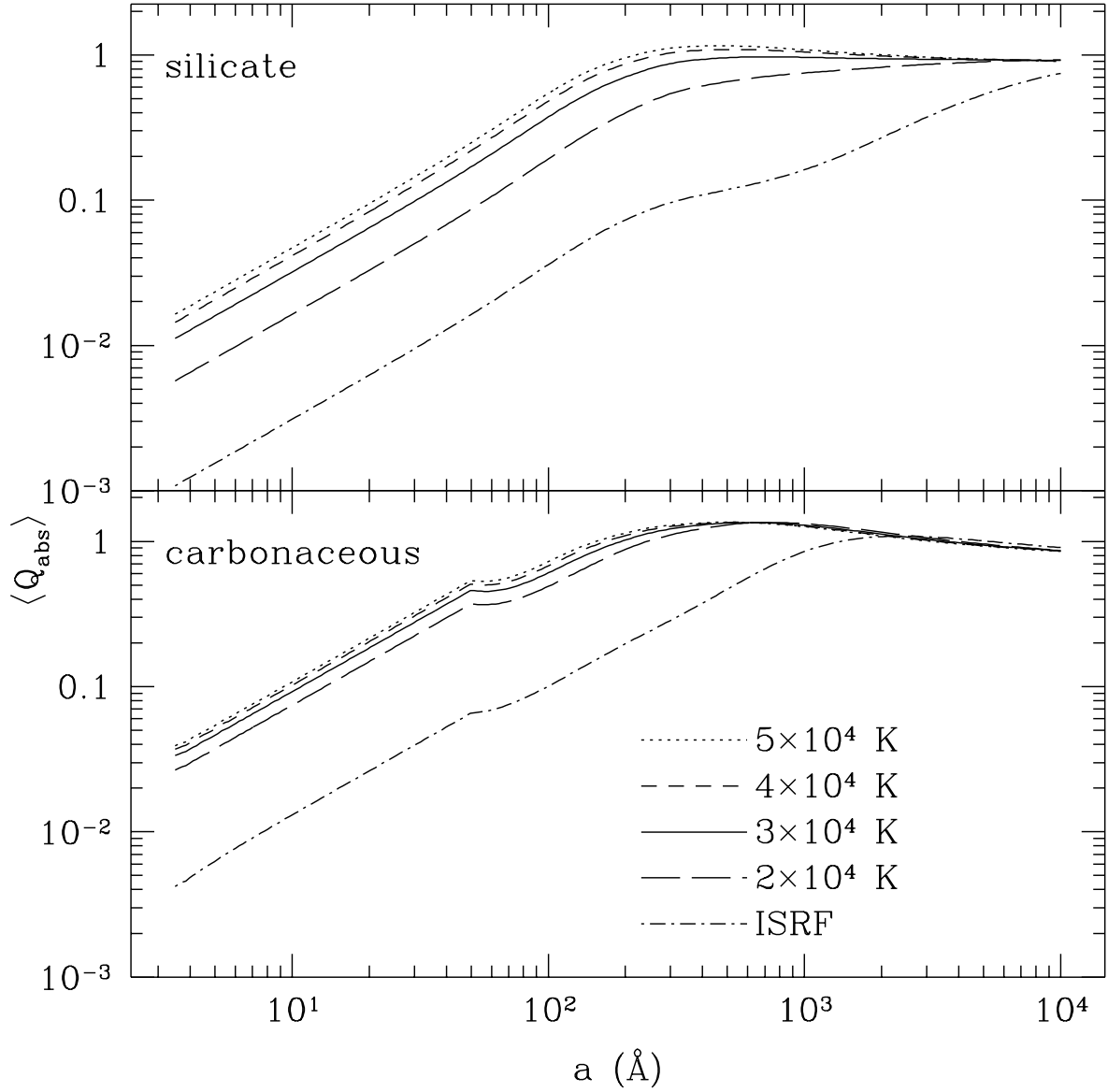}
\caption{
\label{fig:qabs}
        Absorption efficiency factors for neutral carbonaceous 
	and silicate
	grains, averaged over the interstellar radiation field (ISRF) 
	and blackbody spectra with indicated color temperatures (cut off 
        at $13.6 \eV$).  The kink
	at $a=50 \Angstrom$ results from the Li \& Draine (2000) 
	prescription for blending PAH and graphite optical properties.
        }
\end{figure}

\begin{figure}
\epsscale{1.00}
\plotone{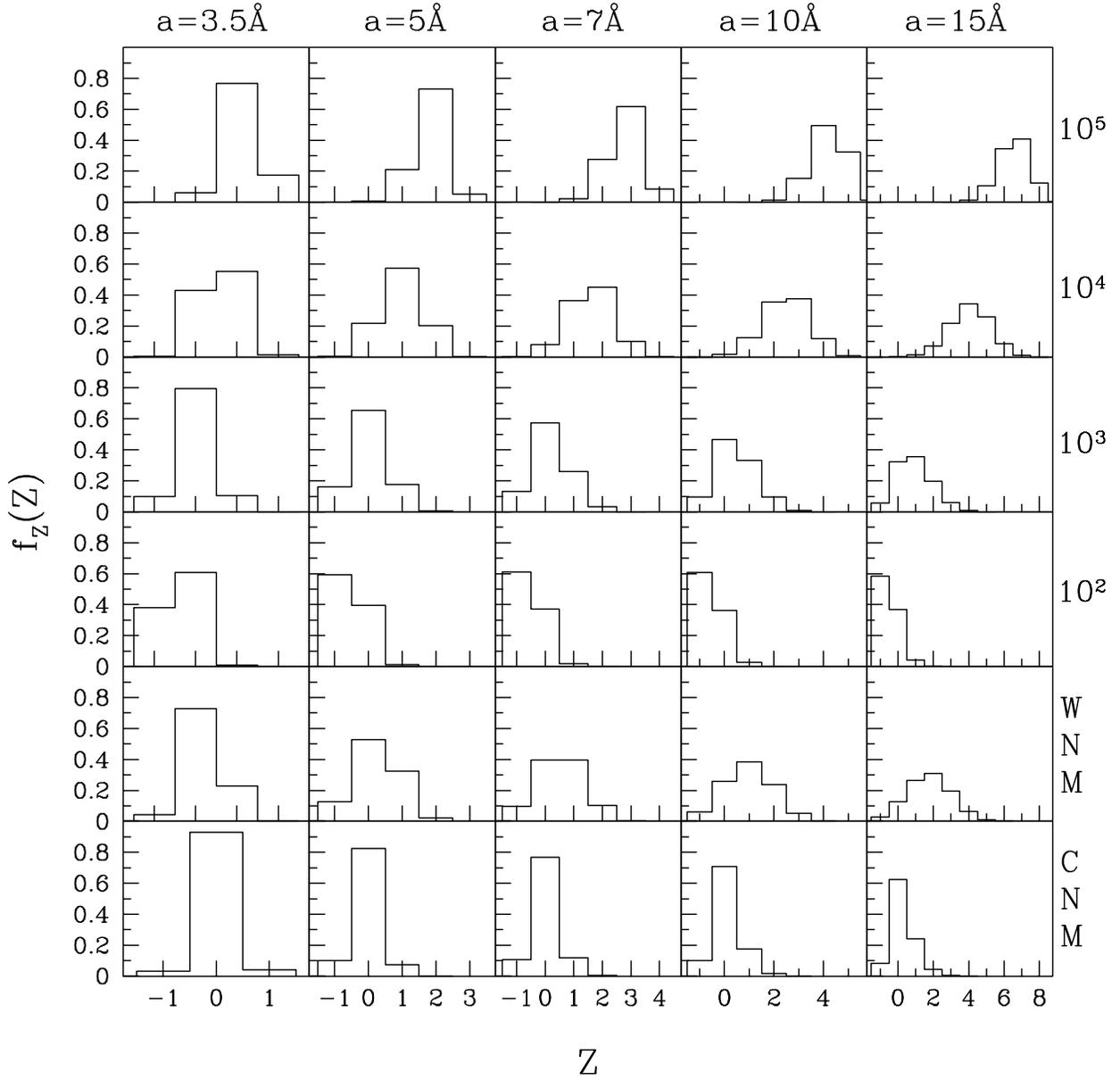}
\caption{
\label{fig:chargedists}
Charge distributions for carbonaceous grains with $a= 3.5$, 5, 7, 10, and
$15 \Angstrom$ and six sets of ambient
conditions.  For the upper four panels, we adopt a blackbody radiation 
spectrum (cut off at $13.6 \eV$) with $\Tc = 3 \times 10^4 \K$ and
$T=10^3 \K$; the values of $G \sqrt{T}/n_e$, in $\K^{1/2} \cm^3$ are 
indicated.  The bottom two panels are for the warm neutral medium (WNM) and
cold neutral medium (CNM). 
        }
\end{figure}

\begin{figure}
\epsscale{1.00}
\plotone{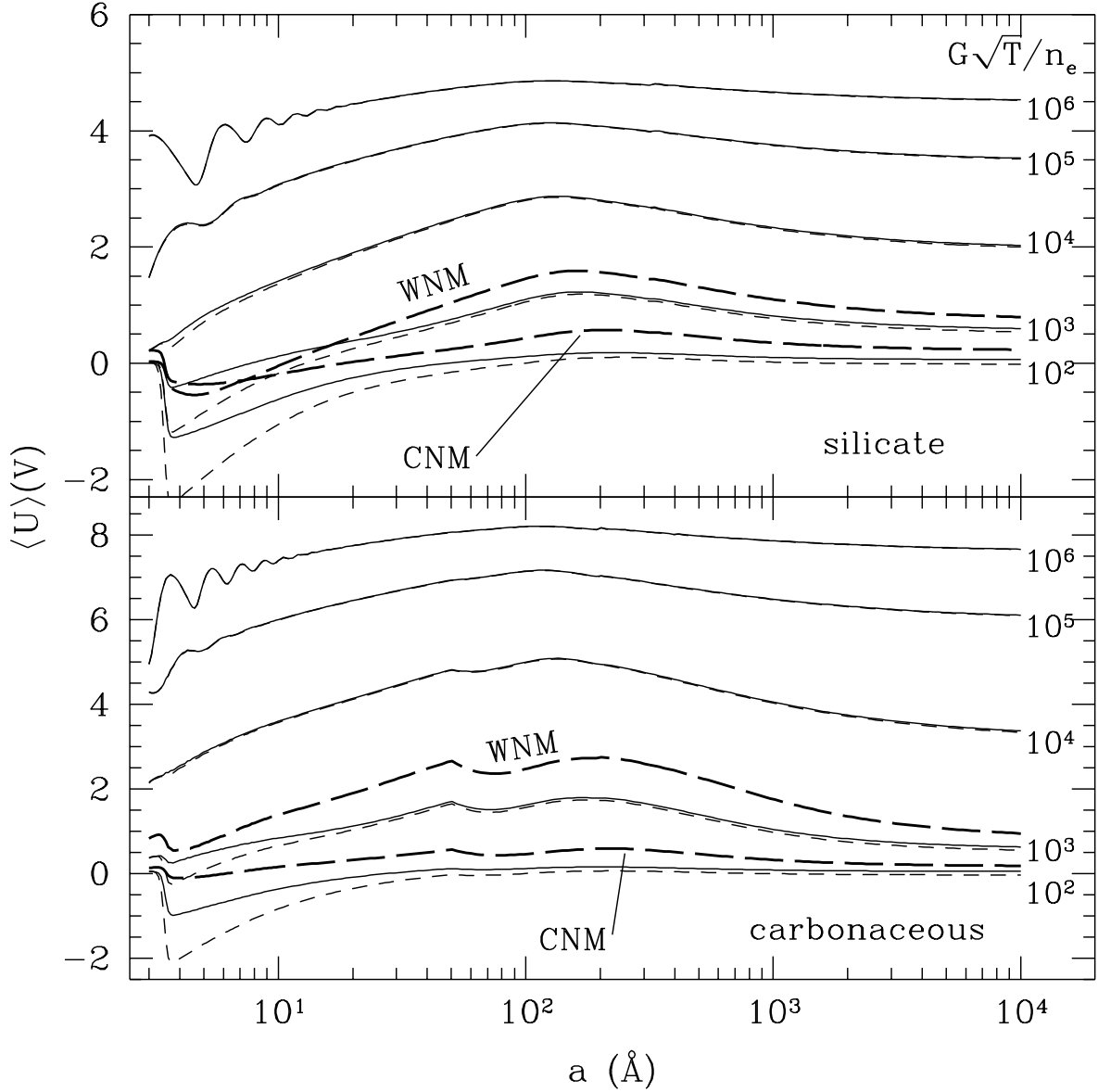}
\caption{
\label{fig:potential}
        Average electrostatic potential $\left<U\right>$ for a blackbody
radiation field with $\Tc = 3 \times 10^4 \K$ (cut off at $13.6 \eV$).  
The top (bottom) panel is for silicate
(carbonaceous) grains; curves labelled CNM (WNM) are for cold (warm)
neutral media; solid (short-dashed) lines indicate $T = 100 \K$ ($1000 \K$).
For the $T=100 \K$ and $1000 \K$ curves, values of $G \sqrt{T} / n_e$ are
indicated in ${\K}^{1/2}{\cm}^3$.
        }
\end{figure}

\begin{figure}
\epsscale{1.00}
\plotone{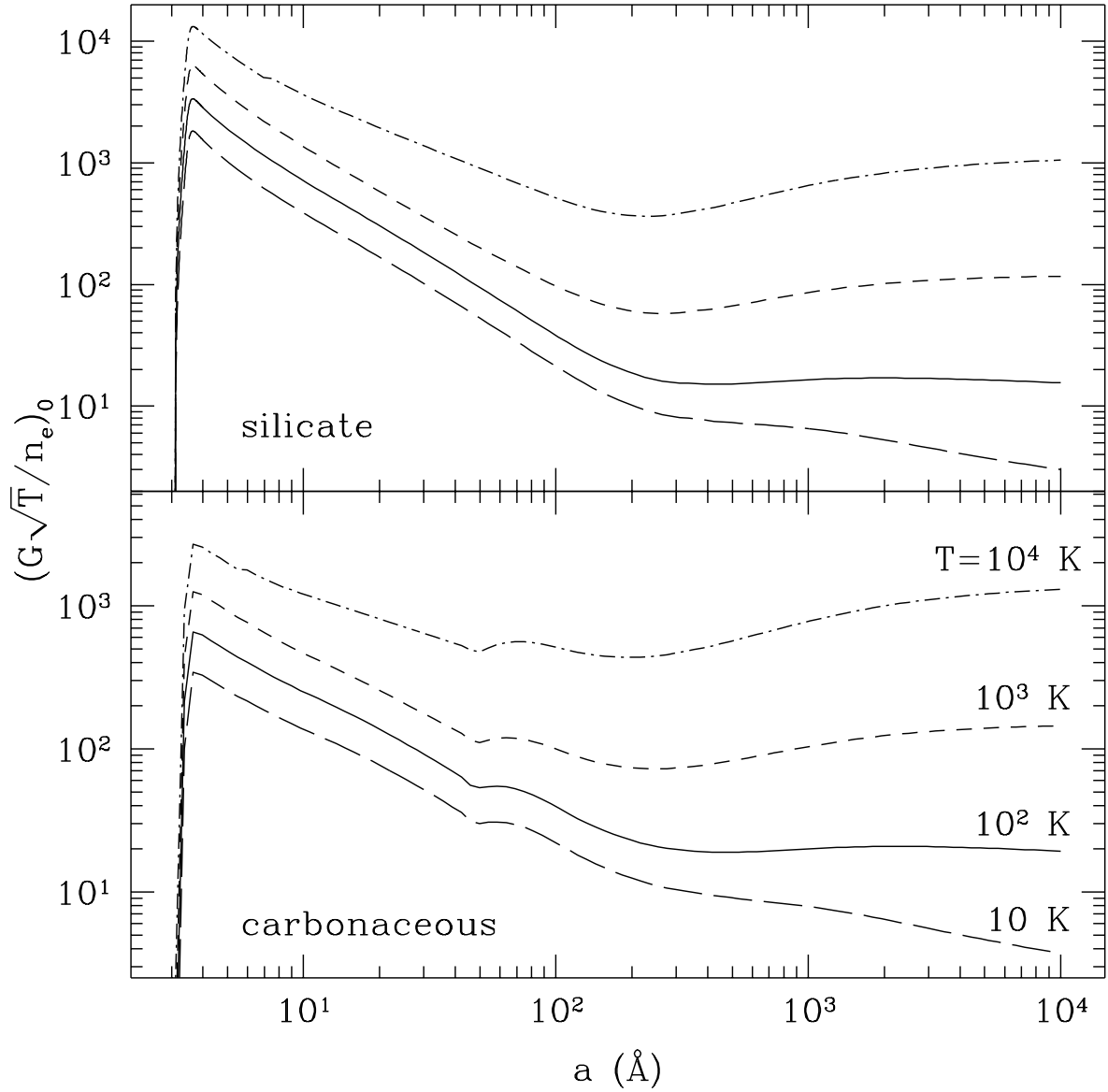}
\caption{
\label{fig:critpar}
	The value of the charging parameter for which $\langle Z \rangle = 0$,
for carbonaceous and silicate grains, a blackbody radiation field with 
$\Tc = 3 \times 10^4 \K$ (cut off at $13.6 \eV$), 
and various gas temperatures, as labeled.
        }
\end{figure}

\begin{figure}
\epsscale{1.00}
\plotone{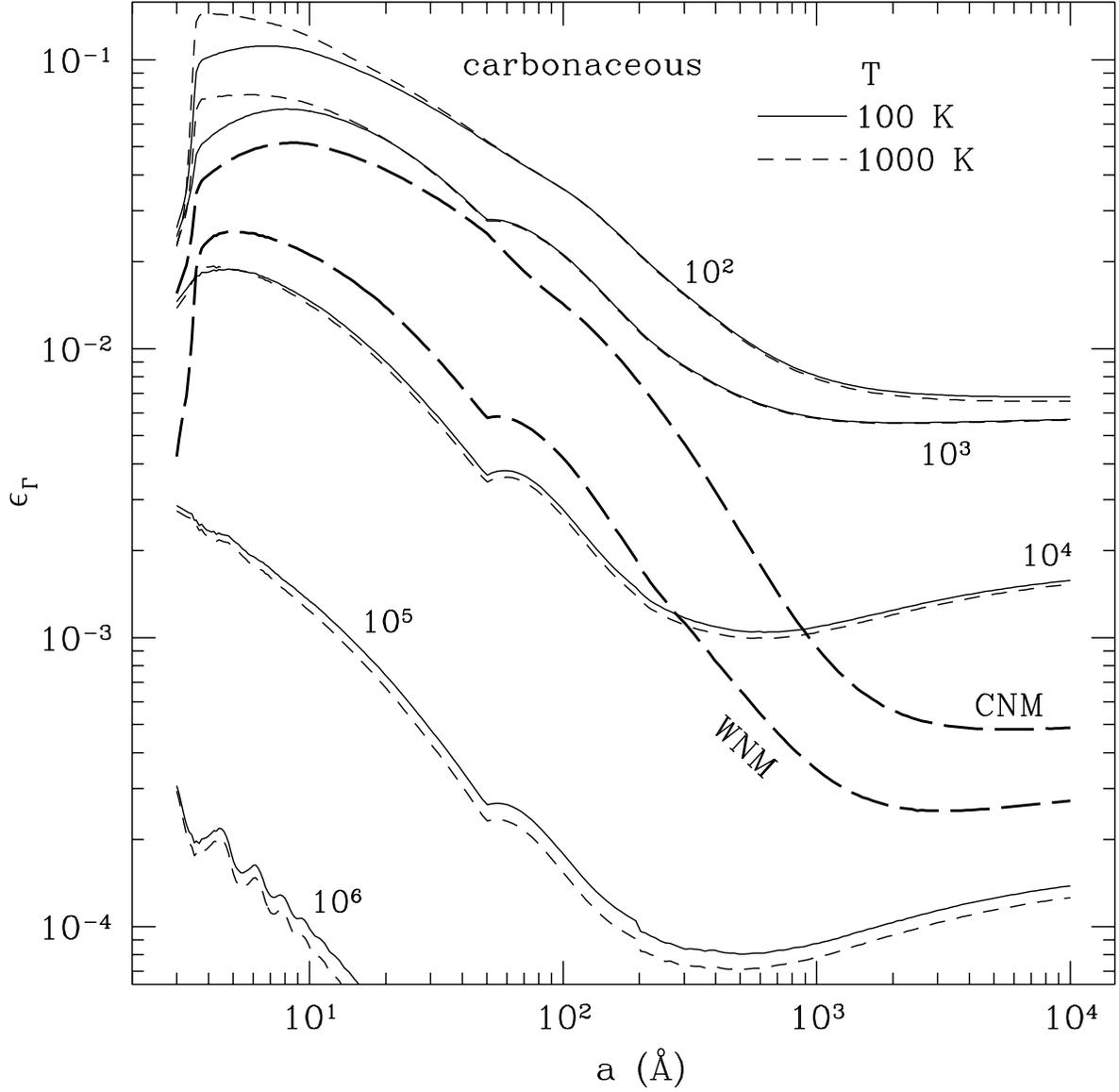}
\caption{
\label{fig:gammapegra}
	Gas heating efficiency for carbonaceous grains
 for two gas temperatures: $100 \K$ (solid) and
$10^3 \K$ (short-dashed), four values of $G \sqrt{T} / n_e$, as labelled,
and $\Tc = 3 \times 10^4 \K$.  (The radiation is cut off at $13.6 \eV$.)
Curves labelled CNM (WNM) are for cold (warm) neutral media and the ISRF.
        }
\end{figure}

\begin{figure}
\epsscale{1.00}
\plotone{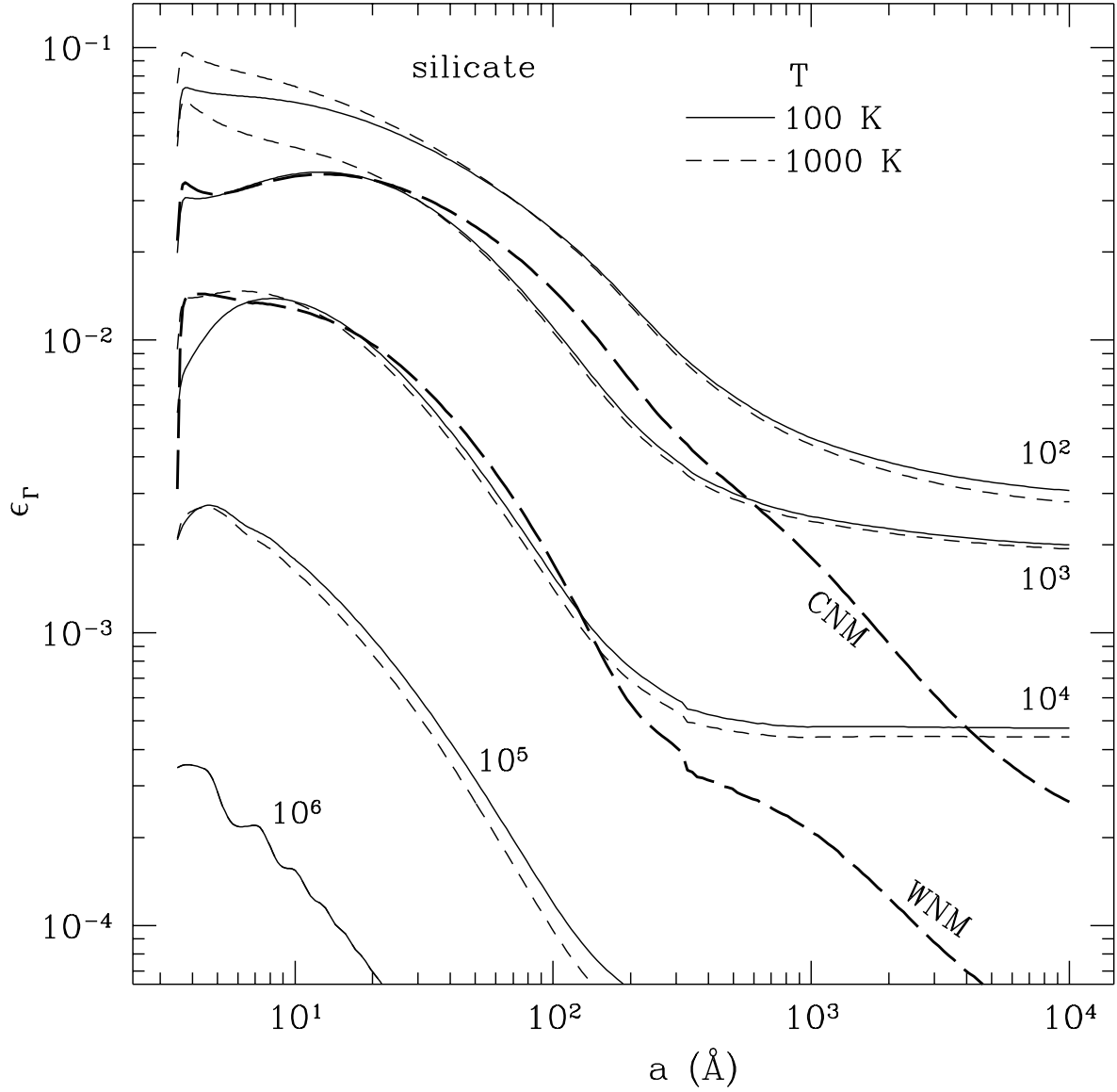}
\caption{
\label{fig:gammapesil}
	Same as Figure \ref{fig:gammapegra}, but for silicate grains.  The 
curve for $T=1000 \K$ and $G \sqrt{T}/n_e = 10^6 \K^{1/2} \cm^3$ lies 
entirely below the plot region.
        }
\end{figure}

\begin{figure}
\epsscale{1.00}
\plotone{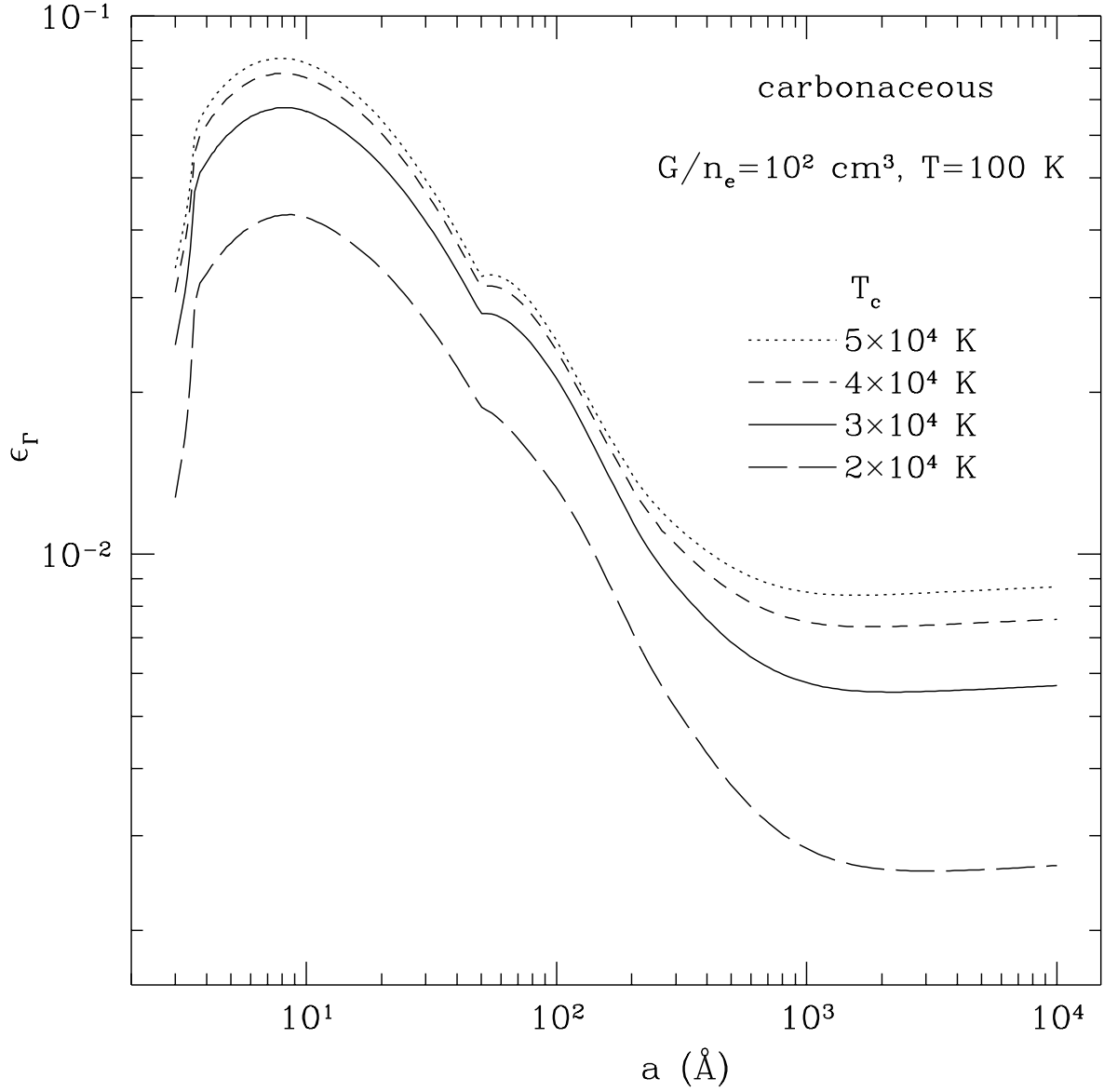}
\caption{
\label{fig:gammapetc}
        Gas heating efficiency for carbonaceous grains, $T = 100 \K$, 
$G \sqrt{T} / n_e = {10}^3 {\K}^{1/2}{\cm}^3$, and four 
values of $\Tc$ as indicated.  (The radiation is cut off at $13.6 \eV$.)
        }
\end{figure}

\begin{figure}
\epsscale{1.00}
\plotone{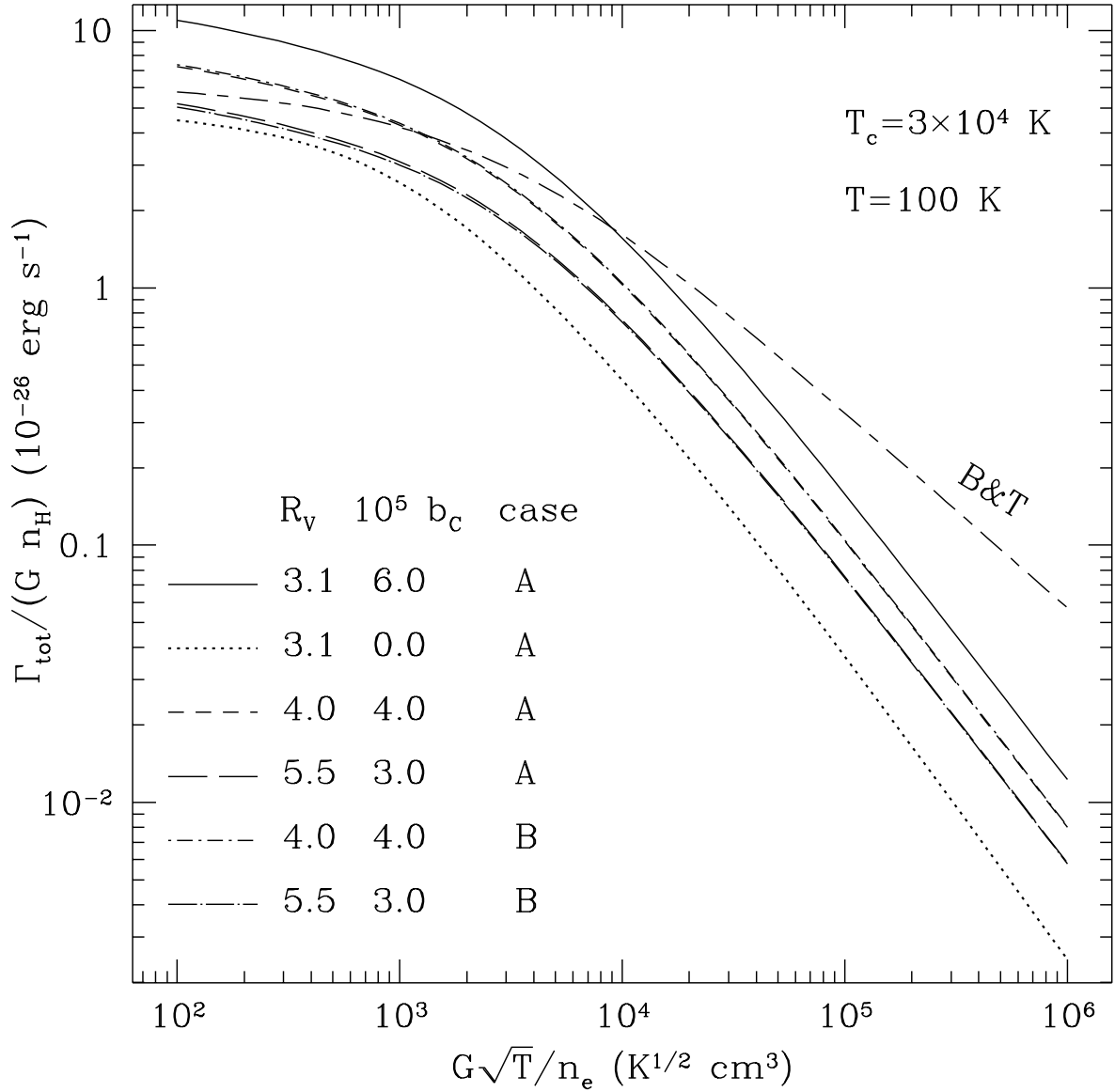}
\caption{
\label{fig:gr_dist_htg}
The photoelectric heating rate minus the cooling rate due to gas phase ions
and electrons colliding with and sticking to the grains; for a blackbody  
radiation spectrum (cut off at $13.6 \eV$) with $\Tc = 3 \times 10^4 \K$,  
gas temperature $T=100 \K$,
and various grain size distributions from Weingartner \& Draine (2000), as
indicated.  For comparison, we also display the Bakes
\& Tielens (1994) result (B\&T).
        }
\end{figure}

\begin{figure}
\epsscale{1.00}
\plotone{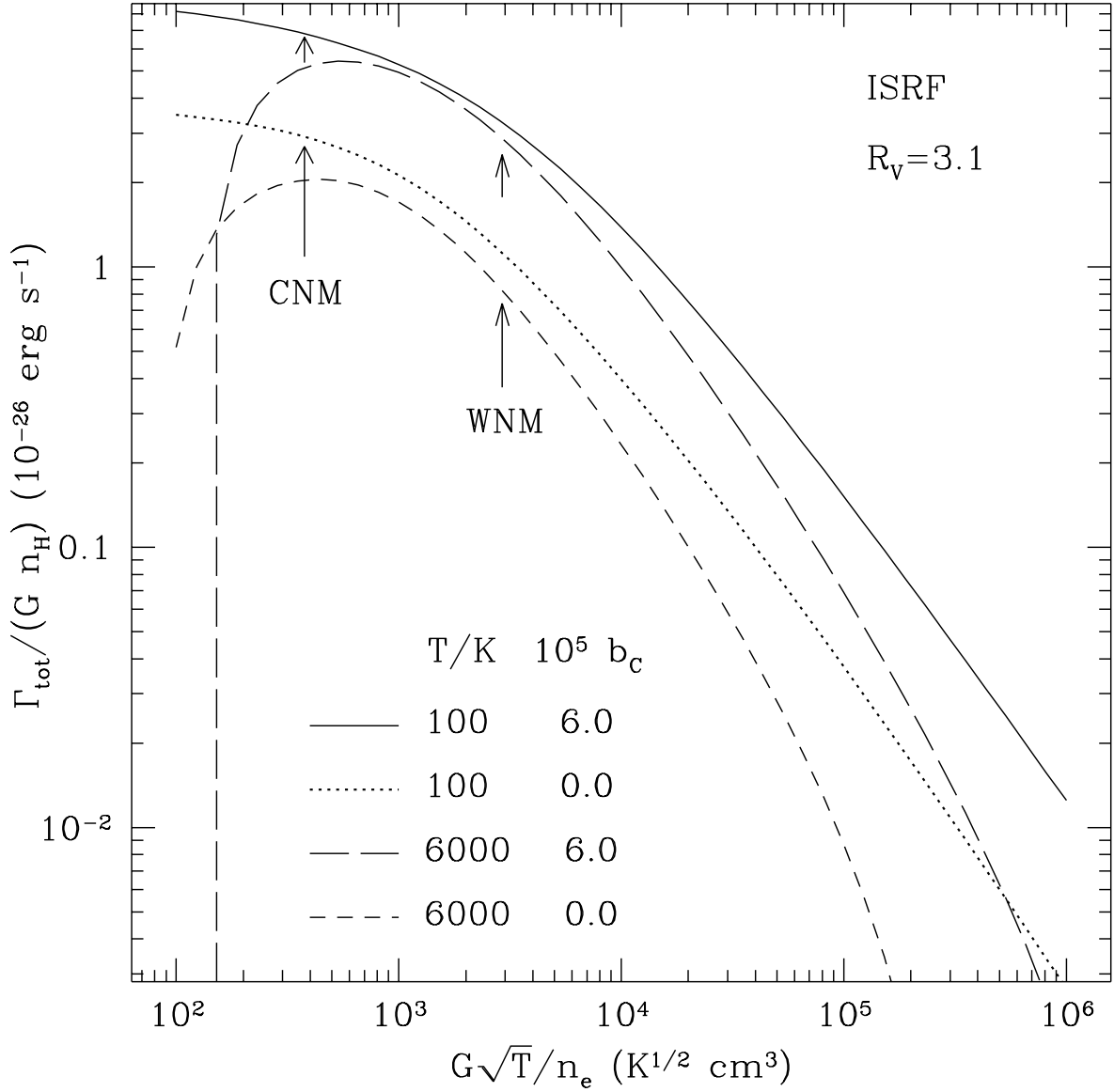}
\caption{
\label{fig:gr_dist_htg_isrf}
Same as Figure \ref{fig:gr_dist_htg}, but for the ISRF of equation 
(\ref{eq:isrf}), $R_V=3.1$, and various values of $b_{\rm C}$ and gas 
temperature $T$.
	}
\end{figure}

\begin{figure}
\epsscale{1.00}
\plotone{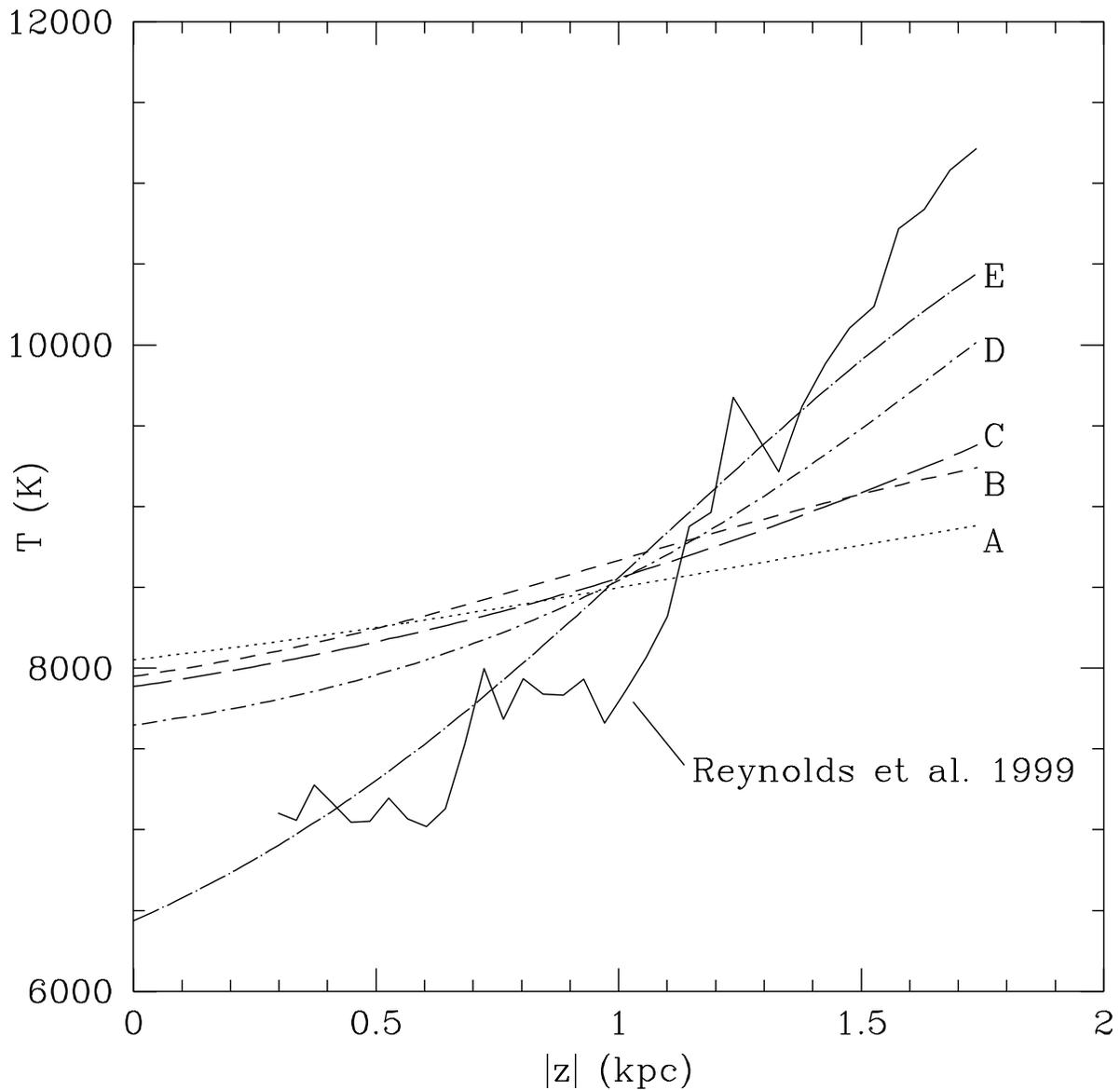}
\caption{
\label{fig:t_wim}
WIM temperature profiles as a function of distance from the 
Galactic midplane $|z|$.  The solid curve was inferred by Reynolds et al.\ 
(1999) from observations of the [N II]/H$\ \alpha$ line intensity ratio; the
other curves are results of calculations including photoelectric heating,
for various sets of parameter values as indicated in Table \ref{tab:wim}.
	}
\end{figure}

\begin{figure}
\epsscale{1.00}
\plotone{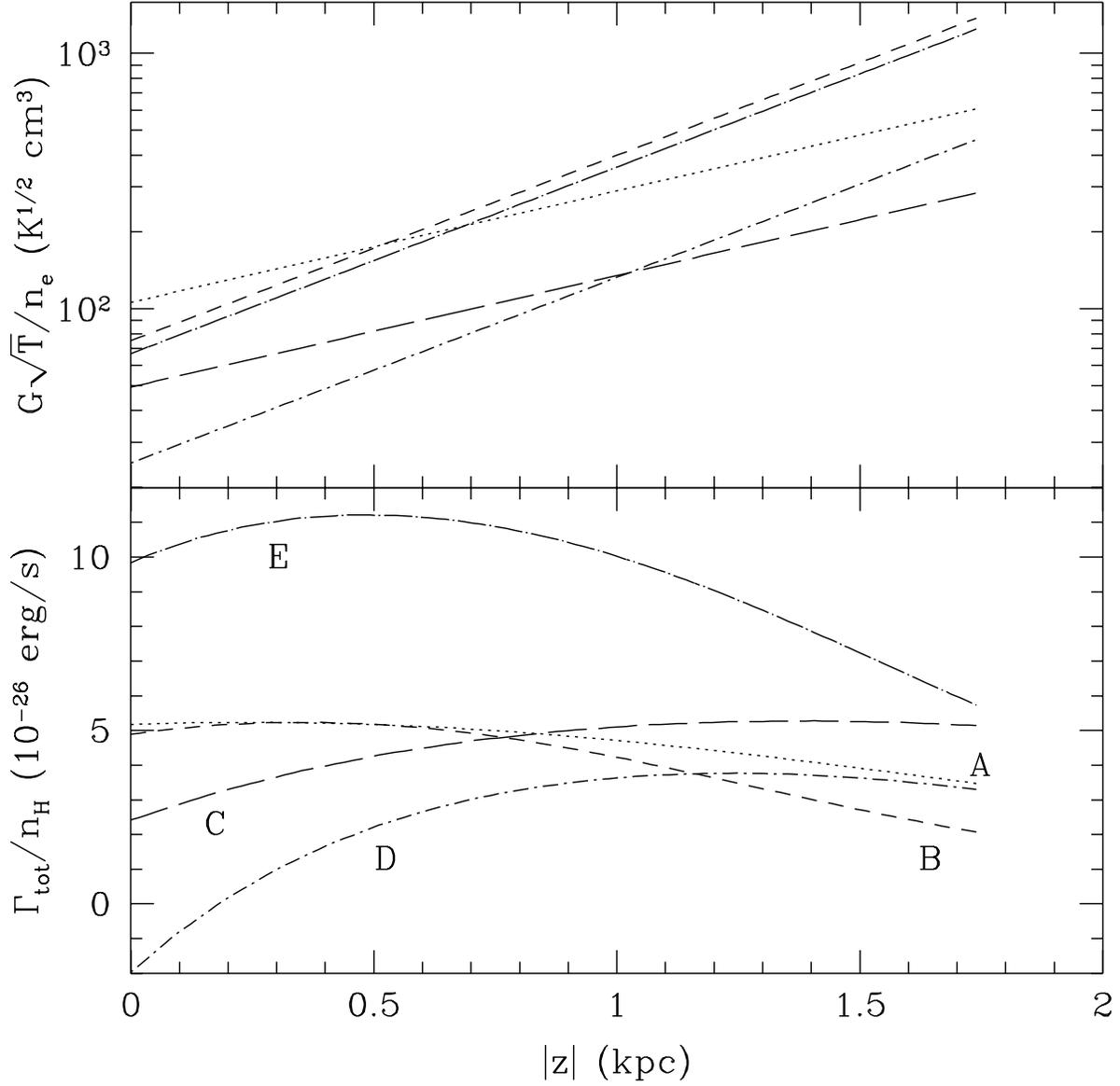}
\caption{
\label{fig:wim2}
Upper panel:  Values of the charging parameter as a function of $|z|$ for the
five cases of Table \ref{tab:wim}.  Lower panel:  The ratio of the net grain
heating rate (i.e. photoelectric heating minus collisional cooling) to the 
gas density as a function of $|z|$, for the five cases of Table \ref{tab:wim}.
	}
\end{figure}

Recent experimental results have been used to obtain new estimates for
electron sticking efficiencies, as a function of grain size, shown in
Figures \ref{fig:s_eattach} and \ref{fig:s_erecomb}.
We have applied our photoemission and collisional charging model to
compute the efficiency with which grains convert absorbed 
radiation into gas heating via the photoelectric effect, 
as a function of grain size (\S 5, 
Figures \ref{fig:gammapegra} -- \ref{fig:gammapetc}).  Bakes \& Tielens
(1994) derived photoelectric heating rates for H~I regions by integrating
their heating efficiencies over an MRN size distribution with the lower 
cutoff size for the graphite population extended to $3.5 \Angstrom$.  
We integrate the net gas heating rate over the size distributions of 
Weingartner \& Draine (2000) (Figures \ref{fig:gr_dist_htg} and 
\ref{fig:gr_dist_htg_isrf}) and we provide fitting functions for the 
resulting total 
photoelectric heating and recombination cooling rates (\S 6).  
In a separate paper, we will examine the consequences of these revised 
rates for the thermal structure of the ISM.  

Finally, we have applied our model to estimate rates for photoelectric
heating by dust grains in H~II regions (\S 7).
We find that photoelectric heating by dust can be important in H~II regions
when $G/\nH \gtsim 1\cm^3$.  Photoelectric heating in the warm ionized medium 
(WIM) might explain the observed variations in emission line intensity 
ratios with distance from the Galactic midplane, if the abundance of
ultrasmall grains is large enough.  The required abundance is equal to that
adopted by Li \& Draine (2000) to account for the diffuse galactic 
infrared emission.

We have investigated a range of grain sizes, gas temperatures, radiation
field color temperatures, and ratios $G/n_e$ of 
radiation intensity to electron density, and selected results have been
presented.  Interested readers can find
a FORTRAN routine that implements the heating
and cooling approximations of equations (\ref{eqn:totheatfit}) and
(\ref{eqn:totcoolfit}) on the World Wide Web at 
www.cita.utoronto.ca/$\sim$weingart.

\acknowledgements
This research was supported in part by NSF grant
AST-9619429 and by NSF Graduate and International Research Fellowships to JCW.
We are grateful to P. A. M. van Hoof for helpful comments on the 
manuscript, L. M. Haffner for providing us with an electronic version 
of the  Reynolds et al.\ (1999) $T(z)$ data, and  
R. H. Lupton for the availability of the SM plotting 
package.


\clearpage


\clearpage

\begin{deluxetable}{ccc}
\tablecaption{Blackbody Components of the ISRF (Mathis et al.\ 1983)
\label{tab:ISRF}}
\tablehead{
\colhead{$i$}&
\colhead{$w_i$}&
\colhead{$T_i / {\rm K}$}
}
\startdata
1	&$10^{-14}$	 		&7500		\\
2	&$1.65 \times 10^{-13}$  	&4000		\\
3	&$4 \times 10^{-13}$		&3000		\\
\enddata
\end{deluxetable}

\begin{deluxetable}{ccccccccccccc}
\tablecaption{Photoelectric Heating Parameters\tablenotemark{a}
\label{tab:totheat}}
\tablehead{
\colhead{$R_V$}&
\colhead{$b_{\rm C}$}&
\colhead{case}&
\colhead{rad field\tablenotemark{b}}&
\colhead{$C_0$}&
\colhead{$C_1$}&
\colhead{$C_2$}&
\colhead{$C_3$}&
\colhead{$C_4$}&
\colhead{$C_5$}&
\colhead{$C_6$}&
\colhead{err}&
\colhead{$h_s$\tablenotemark{c}}
}
\startdata
3.1 &0.0 &A &B0   &5.56 &$1.82 \times 10^{-2}$ &0.00492 &0.03368 &0.557 
&0.666 &0.532 &0.15 &0.75 \\
3.1 &2.0 &A &B0   &3.41 &$4.27$                &0.03700 &0.00569 &0.102 
&0.494 &0.669 &0.19 &0.89 \\
3.1 &4.0 &A &B0  &10.59 &$4.05 \times 10^{-3}$ &0.01234 &0.01391 &0.808 
&0.580 &0.573 &0.19 &0.94 \\
3.1 &6.0 &A &B0   &3.66 &$7.66$                &0.00661 &0.01552 &0.094 
&0.697 &0.482 &0.20 &0.96 \\
4.0 &0.0 &A &B0   &4.30 &$1.79 \times 10^{-4}$ &0.00572 &0.02488 &1.026 
&0.701 &0.505 &0.15 &0.77 \\
4.0 &2.0 &A &B0   &5.03 &$2.27$                &0.07441 &0.00345 &0.140 
&0.412 &0.737 &0.18 &0.92 \\
4.0 &4.0 &A &B0   &3.28 &$4.44$                &0.00786 &0.01219 &0.102 
&0.686 &0.500 &0.20 &0.96 \\
5.5 &0.0 &A &B0   &3.67 &$2.42 \times 10^{-3}$ &0.07034 &0.00360 &0.725 
&0.365 &0.814 &0.15 &0.77 \\
5.5 &1.0 &A &B0   &5.45 &$2.44 \times 10^{-2}$ &0.09594 &0.00225 &0.534 
&0.418 &0.758 &0.18 &0.89 \\
5.5 &2.0 &A &B0   &4.87 &$7.28 \times 10^{-3}$ &0.00812 &0.01969 &0.656 
&0.620 &0.534 &0.19 &0.94 \\
5.5 &3.0 &A &B0   &5.93 &$9.26 \times 10^{-2}$ &0.00675 &0.01648 &0.417
&0.683 &0.489 &0.20 &0.96 \\
4.0 &0.0 &B &B0   &4.85 &$6.59 \times 10^{-2}$ &0.00437 &0.04234 &0.415 
&0.664 &0.508 &0.16 &0.86 \\
4.0 &2.0 &B &B0   &5.97 &$0.287$               &0.00736 &0.02065 &0.302 
&0.637 &0.528 &0.18 &0.93 \\
4.0 &4.0 &B &B0   &7.97 &$0.175$               &0.00277 &0.03820 &0.382 
&0.771 &0.401 &0.19 &0.96 \\
5.5 &0.0 &B &B0   &4.04 &$0.653$               &0.11310 &0.00259 &0.198 
&0.348 &0.809 &0.18 &0.89 \\
5.5 &1.0 &B &B0   &3.38 &$0.608$               &0.00396 &0.02568 &0.192 
&0.752 &0.439 &0.18 &0.93 \\
5.5 &2.0 &B &B0   &1.10 &$3.30$                &0.00941 &0.01431 &0.087 
&0.643 &0.524 &0.20 &0.95 \\
5.5 &3.0 &B &B0   &3.47 &$1.77$                &0.00494 &0.01953 &0.140
&0.723 &0.456 &0.20 &0.97 \\
3.1 &0.0 &A &ISRF &4.35 &$2.63 \times 10^{-5}$ &0.00242 &0.03003 &1.235 
&0.827 &0.399 &0.15 &0.73 \\
3.1 &2.0 &A &ISRF &4.87 &$0.948$               &0.02576 &0.00766 &0.188 
&0.490 &0.650 &0.18 &0.88 \\
3.1 &4.0 &A &ISRF &7.41 &$0.772$               &0.03895 &0.00606 &0.239 
&0.408 &0.708 &0.19 &0.94 \\
3.1 &6.0 &A &ISRF &9.30 &$0.248$               &0.00697 &0.01848 &0.365 
&0.633 &0.509 &0.20 &0.96 \\
\enddata
\tablenotetext{a}{See eq (\ref{eqn:totheatfit}).}
\tablenotetext{b}{B0 refers to a blackbody spectrum with 
$\Tc = 3 \times 10^4 \K$ and the ISRF of Mathis et al.\ (1983) is defined
in eq (\ref{eq:isrf}).}
\tablenotetext{c}{Fraction of the total heating due to grains with 
$a < 100 \Angstrom$, for $G \sqrt{T} / n_e = 10^2 \K^{1/2} \cm^3$ and
$T = 100 \K$.}
\end{deluxetable}

\begin{deluxetable}{cccccccccc}
\tablecaption{Collisional Cooling Parameters\tablenotemark{a}
\label{tab:coolpars}}
\tablehead{
\colhead{$R_V$}&
\colhead{$b_{\rm C}$}&
\colhead{case}&
\colhead{rad field\tablenotemark{b}}&
\colhead{$D_0$}&
\colhead{$D_1$}&
\colhead{$D_2$}&
\colhead{$D_3$}&
\colhead{$D_4$}&
\colhead{err}
}
\startdata
3.1 &0.0 &A &B0   &0.4800 &1.783 &-7.617 &1.655 &0.06326 &0.17 \\
3.1 &2.0 &A &B0   &0.5322 &1.325 &-6.047 &1.435 &0.05378 &0.14 \\
3.1 &4.0 &A &B0   &0.4336 &2.108 &-7.359 &1.736 &0.06502 &0.14 \\
3.1 &6.0 &A &B0   &0.4270 &2.120 &-7.301 &1.786 &0.06731 &0.15 \\ 
4.0 &0.0 &A &B0   &0.5332 &1.379 &-7.125 &1.504 &0.05732 &0.17 \\ 
4.0 &2.0 &A &B0   &0.4425 &2.019 &-7.755 &1.712 &0.06424 &0.14 \\
4.0 &4.0 &A &B0   &0.4847 &1.493 &-6.109 &1.528 &0.05751 &0.13 \\
5.5 &0.0 &A &B0   &0.4184 &2.129 &-9.082 &1.818 &0.06943 &0.16 \\
5.5 &1.0 &A &B0   &0.4981 &1.510 &-7.237 &1.535 &0.05777 &0.14 \\
5.5 &2.0 &A &B0   &0.4851 &1.501 &-6.775 &1.533 &0.05762 &0.13 \\
5.5 &3.0 &A &B0   &0.4948 &1.442 &-6.377 &1.505 &0.05645 &0.13 \\
4.0 &0.0 &B &B0   &0.4568 &1.876 &-7.810 &1.688 &0.06412 &0.15 \\
4.0 &2.0 &B &B0   &0.4597 &1.594 &-6.675 &1.599 &0.06094 &0.14 \\
4.0 &4.0 &B &B0   &0.5155 &1.340 &-5.791 &1.449 &0.05411 &0.13 \\ 
5.5 &0.0 &B &B0   &0.4360 &1.996 &-8.372 &1.735 &0.06572 &0.15 \\
5.5 &1.0 &B &B0   &0.5140 &1.349 &-6.676 &1.472 &0.05536 &0.14 \\
5.5 &2.0 &B &B0   &0.4633 &1.690 &-7.158 &1.607 &0.06055 &0.14 \\
5.5 &3.0 &B &B0   &0.4677 &1.780 &-7.173 &1.615 &0.06027 &0.13 \\
3.1 &0.0 &A &ISRF &0.4291 &2.406 &-8.357 &1.714 &0.06354 &0.15 \\
3.1 &2.0 &A &ISRF &0.5232 &1.678 &-5.942 &1.339 &0.04813 &0.14 \\
3.1 &4.0 &A &ISRF &0.3959 &2.380 &-6.554 &1.575 &0.05674 &0.13 \\
3.1 &6.0 &A &ISRF &0.3632 &2.937 &-7.601 &1.742 &0.06228 &0.12 \\
\enddata
\tablenotetext{a}{See eq (\ref{eqn:totcoolfit}).}
\tablenotetext{b}{B0 refers to a blackbody spectrum with
$\Tc = 3 \times 10^4 \K$ and the ISRF of Mathis et al.\ (1983) is defined
in eq (\ref{eq:isrf}).}
\end{deluxetable}

\begin{deluxetable}{cccccccccccc}
\tablecolumns{12}
\tablewidth{0pc}  
\tablecaption{Photoelectric Heating Rates for H~II Regions
\label{tab:heatHII}
}
\tablehead{
\colhead{} & \colhead{} & \colhead{} & \colhead{} & \colhead{}
& \multicolumn{3}{c}{$\Gamma_{\rm pe}/G \nH$ \tablenotemark{a,{\rm b}}} 
& \colhead{}
& \multicolumn{3}{c}{$\Gamma_{\rm pe}/G \nH$ \tablenotemark{a,{\rm c}}} \\
\cline{6-8} \cline{10-12} \\
\colhead{$R_V$\tablenotemark{d}}&
\colhead{$10^5 b_{\rm C}$\tablenotemark{e}}&
\colhead{case}&
\colhead{$\Tc$ \tablenotemark{f}}&
\colhead{$G/\nH =$}&
\colhead{$0.1$ \tablenotemark{g}}&
\colhead{$1.0$}&
\colhead{$10.$}&
\colhead{}&
\colhead{$0.1$}&
\colhead{$1.0$}&
\colhead{$10.$}
}
\startdata
$3.1$ &$0.0$  &A &$3.5$ & &$1.03$ &$0.83$ &$0.40$ & &$3.58$ &$2.91$ &$1.25$ \\
$3.1$ &$6.0$  &A &$3.5$ & &$2.35$ &$2.06$ &$1.16$ & &$8.15$ &$7.07$ &$3.59$ \\
$4.0$ &$4.0$  &A &$3.5$ & &$1.54$ &$1.36$ &$0.76$ & &$5.37$ &$4.66$ &$2.35$ \\
$5.5$ &$3.0$  &A &$3.5$ & &$1.09$ &$0.97$ &$0.55$ & &$3.83$ &$3.34$ &$1.68$ \\
$4.0$ &$4.0$  &B &$3.5$ & &$1.57$ &$1.38$ &$0.77$ & &$5.49$ &$4.76$ &$2.38$ \\
$5.5$ &$3.0$  &B &$3.5$ & &$1.06$ &$0.94$ &$0.54$ & &$3.74$ &$3.28$ &$1.64$ \\
$3.1$ &$0.0$  &A &$4.5$ & &$1.13$ &$0.91$ &$0.43$ & &$5.93$ &$4.73$ &$1.80$ \\
$3.1$ &$6.0$  &A &$4.5$ & &$2.57$ &$2.26$ &$1.24$ & &$13.1$ &$11.1$ &$5.02$ \\
$4.0$ &$4.0$  &A &$4.5$ & &$1.68$ &$1.48$ &$0.82$ & &$8.62$ &$7.31$ &$3.25$ \\
$5.5$ &$3.0$  &A &$4.5$ & &$1.19$ &$1.06$ &$0.59$ & &$6.13$ &$5.21$ &$2.31$ \\
$4.0$ &$4.0$  &B &$4.5$ & &$1.72$ &$1.51$ &$0.83$ & &$8.81$ &$7.46$ &$3.30$ \\
$5.5$ &$3.0$  &B &$4.5$ & &$1.16$ &$1.03$ &$0.58$ & &$5.96$ &$5.08$ &$2.26$ \\
\enddata
\tablenotetext{a}{$10^{-25} \erg \, {\rm s}^{-1}$}
\tablenotetext{b}{Blackbody spectrum cut off at $13.6 \eV$}
\tablenotetext{c}{Full blackbody spectrum}
\tablenotetext{d}{$R_V = A_V/E_{B-V}$, ratio of visual extinction to
reddening}
\tablenotetext{e}{C abundance in very small grain population}
\tablenotetext{f}{$10^4 \K$}
\tablenotetext{g}{Value for $G/\nH$, in $\cm^3$}
\end{deluxetable}

\begin{deluxetable}{cccccccccccc}
\tablecolumns{12}
\tablewidth{0pc}
\tablecaption{Recombination Cooling Rates for H~II Regions
\label{tab:coolHII}
}
\tablehead{
\colhead{} & \colhead{} & \colhead{} & \colhead{} & \colhead{}
& \multicolumn{3}{c}{$\Lambda /G \nH$ \tablenotemark{a,{\rm b}}}
& \colhead{}
& \multicolumn{3}{c}{$\Lambda /G \nH$ \tablenotemark{a,{\rm c}}} \\
\cline{6-8} \cline{10-12} \\
\colhead{$R_V$\tablenotemark{d}}&
\colhead{$b_{\rm C}$\tablenotemark{e}}&
\colhead{case}&
\colhead{$\Tc$ \tablenotemark{f}}&
\colhead{$G/\nH =$}&
\colhead{$0.1$ \tablenotemark{g}}&
\colhead{$1.0$}&
\colhead{$10.$}&
\colhead{}&
\colhead{$0.1$}&
\colhead{$1.0$}&
\colhead{$10.$}
}
\startdata
$3.1$ &$0.0$ &A &$3.5$ & &$7.43$ &$1.02$ &$0.22$ & &$7.91$ &$1.36$ &$0.36$ \\
$3.1$ &$4.0$ &A &$3.5$ & &$29.4$ &$3.47$ &$0.64$ & &$30.2$ &$4.17$ &$1.00$ \\
$4.0$ &$2.0$ &A &$3.5$ & &$18.7$ &$2.20$ &$0.41$ & &$19.2$ &$2.66$ &$0.64$ \\
$5.5$ &$1.0$ &A &$3.5$ & &$13.2$ &$1.55$ &$0.29$ & &$13.6$ &$1.87$ &$0.45$ \\
$4.0$ &$2.0$ &B &$3.5$ & &$18.9$ &$2.23$ &$0.42$ & &$19.4$ &$2.69$ &$0.65$ \\
$5.5$ &$1.0$ &B &$3.5$ & &$13.1$ &$1.53$ &$0.28$ & &$13.5$ &$1.84$ &$0.44$ \\
$3.1$ &$0.0$ &A &$4.5$ & &$7.46$ &$1.05$ &$0.23$ & &$8.29$ &$1.58$ &$0.44$ \\
$3.1$ &$4.0$ &A &$4.5$ & &$29.4$ &$3.51$ &$0.67$ & &$30.9$ &$4.68$ &$1.21$ \\
$4.0$ &$2.0$ &A &$4.5$ & &$18.7$ &$2.23$ &$0.43$ & &$19.6$ &$2.98$ &$0.78$ \\
$5.5$ &$1.0$ &A &$4.5$ & &$13.2$ &$1.57$ &$0.30$ & &$13.9$ &$2.10$ &$0.55$ \\
$4.0$ &$2.0$ &B &$4.5$ & &$18.9$ &$2.26$ &$0.43$ & &$19.9$ &$3.02$ &$0.79$ \\
$5.5$ &$1.0$ &B &$4.5$ & &$13.1$ &$1.55$ &$0.29$ & &$13.8$ &$2.06$ &$0.54$ \\
\enddata
\tablenotetext{a}{$10^{-25} \erg \, {\rm s}^{-1}$}
\tablenotetext{b}{Blackbody spectrum cut off at $13.6 \eV$}
\tablenotetext{c}{Full blackbody spectrum}
\tablenotetext{d}{$R_V = A_V/E_{B-V}$, ratio of visual extinction to 
reddening}
\tablenotetext{e}{C abundance in very small grain population}
\tablenotetext{f}{$10^4 \K$}
\tablenotetext{g}{Value for $G/\nH$, in $\cm^3$}
\end{deluxetable}

\begin{deluxetable}{ccccc}
\tablecaption{WIM Heating Examples
\label{tab:wim}}
\tablehead{
\colhead{Case}&
\colhead{Grain Size Distribution}&
\colhead{WIM Filling Factor}&
\colhead{
$\Gamma_{\rm pi}/n_e^2T_4^{-0.8}$}&
\colhead{$G$}
\\
\colhead{}&
\colhead{}&
\colhead{}&
\colhead{$10^{-24} \erg \cm^3 \s^{-1}$}&
\colhead{}
}
\startdata
A &$R_V=3.1$, $b_{\rm C}=0.0$  &0.2 &1.5 &0.3  \\
B &$R_V=3.1$, $b_{\rm C}=0.0$  &$0.1 \exp{|z|/750 \, {\rm pc}}$ &1.5 &0.3  \\
C &$R_V=3.1$, $b_{\rm C}=6 \times 10^{-5}$  &0.2 &1.5 &0.14 \\
D &$R_V=3.1$, $b_{\rm C}=6 \times 10^{-5}$  &$0.1 \exp{|z|/750 \, {\rm pc}}$ 
&1.5 &0.1  \\
E &$R_V=3.1$, $b_{\rm C}=6 \times 10^{-5}$  &$0.1 \exp{|z|/750 \, {\rm pc}}$
&0.7 &0.27  \\
\enddata
\end{deluxetable}

\end{document}